\documentclass{aastex6}
\usepackage{graphicx}
\shorttitle{50 MHz -- 50 GHz Flux Density Scale}
\shortauthors{Perley and Butler}
\begin{document}
\title{An Accurate Flux Density Scale from 50 MHz to 50 GHz}
\author{R. A. Perley and B.J. Butler}
\email{RPerley@nrao.edu, BButler@nrao.edu}
\affil{National Radio Astronomy Observatory\\
P.O.Box O, Socorro, NM, 87801}
\slugcomment{Intended for the Astrophysical Journal Supplement}

\begin{abstract}
  The flux density scale of \citet{PB13} is extended downwards to
  $\sim$ 50 MHz by utilizing recent observations with the Karl
  G. Jansky Very Large Array (VLA)\footnote{The National Radio
    Astronomy Observatory is a facility of the National Science
    Foundation operated under cooperative agreement by Associated
    Universities, Inc.} of twenty sources between 220 MHz
  and 48.1 GHz, and legacy VLA observations at 73.8 MHz.  The derived
  spectral flux densities are placed on an absolute scale by utilizing
  the \citet{Baa77} values for Cygnus A (3C405) for frequencies below
  2 GHz, and the Mars-based polynomials for 3C286, 3C295, and 3C196
  from \citet{PB13} above 2 GHz.  Polynomial expressions are presented
  for all 20 sources, with accuracy limited by the primary standards
  to 3 -- 5\% over the entire frequency range.  Corrections to the
  scales proposed by \citet{PB13}, and by \citet{SH12} are given.

  \end{abstract}


\maketitle

\section{Introduction}

The radio astronomy flux density scale has long been based on the
polynomial expressions given in \citet{Baa77} for four `absolute' and
13 `secondary' sources.  The members of the first group -- Cygnus A,
Casseopeia A, Taurus A, and Virgo A -- all have angular extents of
arcminutes and high spectral flux densities, typically in excess of
1000 Jy at $\lambda = $1m.  They are sufficiently strong that their
flux densities can be accurately measured with low-resolution
telescopes with known gains, but their large scale structure makes
them unsuitable for calibration by high-resolution interferometric
arrays.  The members of the second group are much smaller -- typically
10s of arcseconds or less, and weaker, with typical flux densities
less than 50 Jy at $\lambda = 1$m.  The flux densities at low
frequencies of these weaker sources could not be accurately measured
by single antennas of known gain, primarily because of source
confusion, and were determined by taking ratios against the members of
the first group, using higher resolution arrays.  A subset of these
`secondary' sources, comprising the smallest amongst them, has been
extensively utilized for flux density calibration of interferometers
in the meter -- centimeter wavelength range.

The \citet{Baa77} (hereafter Baars77) scale for the compact
(`secondary') sources is nominally valid between 0.4 and 15 GHz with
accuracy of $\sim$5\%.  Prior to $\sim$1990, the great majority of
observing with the VLA was at its two lowest frequency bands -- 1.5
and 5 GHz, so there was little incentive to extend the Baars77 flux
density scale above 15 GHz.  After that time, the 15 and 23 GHz
receivers were replaced with much more sensitive ones, and a new band,
centered at 45 GHz was added.  These additions, and the subsequent
holographic surface adjustments and improvements in observing
methodologies needed to support high frequency VLA observing, resulted
in a greatly increased use of high frequencies on the VLA,
necessitating extension of the Baars77 scale to higher frequencies.
Accurate measurements of a set of small-diameter radio sources by the
VLA resulted in the \citet{PB13} (herafter PB13) scale, which was
placed on an absolute scale via VLA and WMAP observations of the
planet Mars, utilizing a thermophysical model of that planet's
emission.  The PB13 scale is valid between 1 and 50 GHz, with claimed
accuracy of $\sim$1\% over the central frequencies, rising to
$\sim$3\% at the highest and lowest frequencies.  However, a
comparison of VLA and ATCA interferometric observations with Planck
observations of 65 compact sources at 22 and 43 GHz by \citet{Par16}
suggests the PB13 scale is low by $\sim$2.5\% at 28 GHz, and by
$\sim$5.5\% at 43 GHz.

Over the past decade, there has been an upsurge in interest in, and
development of, low-frequency radio astronomy.  However, the Baars77
scale for the compact objects useful for calibration purposes is not
defined below 400 MHz.  Indeed the low-frequency flux density scale
has long been quite uncertain, as summarized in the Baars77 paper.
Table 7 of that paper shows the ratio of their scale for the
`absolute' sources to the scales of \citet{CKL63}, \citet{K64},
\citet{BMW65}, \citet{BH72}, and \citet{W73}.  Variations at the
$\sim$10\% level are seen -- most likely due to the effects of
background source confusion to the low-resolution instruments utilized
at the time.  A useful low-frequency scale has been proposed by
\citet{SH12} (hereafter SH12), valid between 30 and 300 MHz.  This
work is a rationalization of 13 different, but interlinked, flux
density scales.  The SH12 scale adopts the B77 scale above 325 MHz,
and the \citet{RBC73} scale below 325 MHz.  The correction factors
needed to adjust the 13 scales to those adopted for the SH12 scale are
given in their Table 2 -- some are as large as 20\%.  Given the
renewed interest in low-frequency astronomy, placing the low frequency
flux density scale on a firm footing, using modern telescopes, is a
worthy endeavor.

Accurately placing the low-frequency flux density scale onto an
absolute standard requires a highly linear, high-resolution array,
preferably comprised of high-gain elements, capable of cleanly
separating the proposed calibration targets from surrounding emission.
As demonstrated by PB13, the VLA is easily capable of determining
ratios between proposed calibration sources to $\sim$1\% accuracy at
its low frequency bands.  Placing the results on an absolute scale
requires a highly linear correlator, as the only suitable `absolute'
reference source over the .02 -- 2 GHz range is Cygnus A, whose flux
density exceeds by more than two orders of magnitude those of the
proposed secondary qcalibrators.  The original Very Large Array was
ill-suited to this task, as its digital correlator was not
sufficiently linear over the large range of correlation coefficients
to directly bootstrap the standard calibrators to Cygnus
A\footnote{For Cygnus A, the correlation coefficient at 300 MHz
  through 1500 MHz is over 0.7 -- which is well beyond the linear
  range for the VLA's original 3-level correlator}.  This situation
changed in 2012, with the commissioning of the Jansky Very Large
Array, and its new correlator, which uses 4-bit (16 level)
correlation.  The result of this was that the response to the absolute
standard source Cygnus A can now be safely linked to weaker -- and
smaller -- radio sources suitable for calibration for interferometers.

Most of the previous work on the flux density scales has been limited
to northern sources, with few southern sources reliably linked to the
absolute standards.  In view of the development of numerous southern
hemisphere low-frequency arrays, it is important to rectify this
situation by inclusion of southern sources.      

In this paper, we propose a single flux density scale, valid from 50
MHz through 50 GHz, based on new observations with the VLA of 19
proposed calibrator sources, over the frequency range 230 MHz to 48
GHz, located in both hemispheres, along with `legacy' observations of
13 sources with the VLA at 73.8 MHz.

\section{Observations}

The results presented here are taken from a more general instrumental
program whose primary goal is to determine and track the VLA's
instrumental parameters over all nine frequency bands.  The source
lists were drawn from the Baars77, PB13 and SH12 papers, with six
additional southern hemisphere objects added to assist in extending
the northern flux calibrator grid to the south.  Added to these were
the four `absolute' Baars77 objects, as a primary goal of these
observations was to link the proposed flux density scale to
these. Three additional far northern calibrator sources were added to
improve the high frequency calibration, as described below.  Not all
sources were observed at all bands -- the large angular extent of many
of the selected sources makes observing them with the VLA impractical
at high frequencies.

For notational brevity, we will utilize `band codes' to represent the
various receiver systems employed on the VLA.  These codes are defined
in Table~\ref{tab:BandCodes}.

\begin{deluxetable}{ccc}
\tablecaption{Band Codes \label{tab:BandCodes}}
\tablehead{
\colhead{Band Code}&\colhead{Representative}&\colhead{Frequency
  Span}\\
&\colhead{Wavelength}&\colhead{(MHz)}}
\startdata
4\tablenotemark{a}&4m&73.0 -- 74.6\\
P\tablenotemark{a} &90cm&224 -- 480\\
L&20cm &1000 -- 2000\\
S&10cm &2000 -- 4000\\
C& 5cm &4000 -- 8000\\
X& 3cm &8000 -- 12000\\
Ku&2cm &12000 -- 18000\\
K&13mm &18000 -- 26500\\
Ka&9mm &26500 -- 40000\\
Q&7mm  &40000 -- 50000\\
\enddata
\tablenotetext{a}{The VLA's low frequency system now comprises a
  single receiver covering 50 -- 480 MHz and two feeds, covering 50 --
  80, and 224 -- 480 MHz, respectively.  The 90cm observations
  described here are taken with this new system.  The 4m observations
  described here are from the narrowband `legacy' system \citep{K07},
  now disabled.}
\end{deluxetable}

The source list, and the bands utilized for each, are given in Table
\ref{tab:SourceList}

\begin{deluxetable}{ccCCCCCCCCCCc}
\tablecaption{Source List \label{tab:SourceList}}
\tablehead{
\colhead{Name}&\colhead{Alternate Names}&\colhead{4}&\colhead{P}&
\colhead{L}&\colhead{S}&\colhead{C}&\colhead{X}&\colhead{Ku}&
\colhead{K}&\colhead{Ka}&\colhead{Q}&
\colhead{LAS\tablenotemark{a}(arcsec)}}
\startdata
J0133$-$3629&                   &       &\bullet&\bullet&       & & &  & &  & &900\\
J0137+3309  &3C48               &\bullet&\bullet&\bullet&\bullet&\bullet&\bullet&\bullet&\bullet&\bullet&\bullet&1.2\\
J0322$-$3712&Fornax A           &       &\bullet&\bullet&       & & &  & &  & &3000\\
J0437+2940  &3C123              &\bullet&\bullet&\bullet&\bullet&\bullet&\bullet&\bullet&\bullet&\bullet&\bullet&43\\
J0444$-$2809&                   &       &\bullet&\bullet&       & & &  & &  & &120\\
J0519$-$4546&Pictor A           &       &\bullet&\bullet&\bullet& & &  & &  & &480\\
J0521+1638  &3C138              &       &\bullet&\bullet&\bullet&\bullet&\bullet&\bullet&\bullet&\bullet&\bullet&0.65\\
J0534+2200  &3C144,Taurus A,Crab&\bullet&\bullet&\bullet&\bullet&\bullet& &  & &  & &480\\
J0542+4951  &3C147              &\bullet&\bullet&\bullet&\bullet&\bullet&\bullet&\bullet&\bullet&\bullet&\bullet&0.70\\
J0813+4813  &3C196              &\bullet&\bullet&\bullet&\bullet&\bullet&\bullet&\bullet&\bullet&\bullet&\bullet&6.0\\
J0918$-$1205&3C218,Hydra A      &\bullet&\bullet&\bullet&\bullet&\bullet&\bullet&\bullet& &  & &420\\
J1230+1223  &3C274,Virgo A,M87  &\bullet&\bullet&\bullet&\bullet& & &  & &  & &840\\
J1331+3030  &3C286              &\bullet&\bullet&\bullet&\bullet&\bullet&\bullet&\bullet&\bullet&\bullet&\bullet&3.5\\
J1411+5212  &3C295              &\bullet&\bullet&\bullet&\bullet&\bullet&\bullet&\bullet&\bullet&\bullet&\bullet&6.5\\
J1651+0459  &3C348,Hercules A   &       &\bullet&\bullet&\bullet&\bullet&\bullet&  & &  & &195\\
J1720$-$0058&3C353              &\bullet&\bullet&\bullet&\bullet& & &  & &  & &320\\
J1829+4844  &3C380              &\bullet&\bullet&\bullet&\bullet&\bullet&\bullet&\bullet&\bullet&\bullet&\bullet&18\\
J1959+4044  &3C405,Cygnus A     &\bullet&\bullet&\bullet&\bullet&\bullet&\bullet&  & &  & &110\\
J2214$-$1701&3C444              &       &\bullet&\bullet&\bullet&\bullet&\bullet&  & &  & &120\\
J2323+5848  &3C461,Cassiopeia A &\bullet&\bullet&\bullet&\bullet& & &  & &  & &480\\
\enddata
\tablenotetext{a}{LAS = Largest Angular Size}
\end{deluxetable}

The observations were made in five observing sessions on the dates
listed in Table~\ref{tab:Log}, which also gives the time spent in each
session, the array configuration in which the observations were made, and a
summary of the utilized frequency bands.

\begin{deluxetable}{cccccc}
\tablecaption{Observing Log 
\label{tab:Log}}
\tablewidth{0pt} 
\tablehead{\colhead{Date}&\colhead{IAT Start}&
\colhead{LST Start}&\colhead{Duration}&
\colhead{Configuration}& \colhead{Comments}\\
&\colhead{hh:mm}&\colhead{hh:mm}&\colhead{h}&&}
\startdata 
07-08 Mar 1998 &17:45&21:30&24&  A&4-band only\\
04-05 Oct 1998 &12:15&06:00&24&  B&4-band only\\
11-12 Oct 2014 &01:00&19:10&30&  C&\\ 
25-26 Jan 2016 &05:00&06:10&27&DnC&D to C reconfig.\\ 
27 Jan 2016    &01:00&02:10& 5&DnC&D to C reconfig.\\ 
\enddata 
\end{deluxetable}

The first two of these sessions were taken under Project ID AK461,
whose goal was to survey the major 3C sources as part of the
commissioning of the initial `4-band' (73 MHz) receiving system
\citep{K07}.  These data were taken with the original VLA correlator,
with 1.6 MHz bandwidth\footnote{Note that at 73.8 MHz, the extremely
  low efficiency of the VLA's antennas ($\sim 0.15$), combined with
  the extremely high sky temperature ($\sim 8000$K) result in a
  correlation coefficient for Cygnus A of only 5\%, well within the
  safe linear range of the old correlator.  Hence, we are able to
  utilize `legacy' observations for this work.}.  All other data were
taken with the VLA system. The additional observations taken in 2016
were to provide more short spacings for better determination of the
flux densities of the highly extended sources in the target list.
Frequency tuning details are given in Table~\ref{tab:Frequencies}.

\begin{deluxetable}{cccc}
\tabletypesize{\scriptsize} 
\tablecaption{Observing Frequencies 
\label{tab:Frequencies}}
\tablewidth{0pt} 
\tablehead{\colhead{Band Code}&\colhead{Frequency Span}&
\colhead{SPW Width}&\colhead{Frequencies Used}\\
&\colhead{MHz}&\colhead{MHz}&\colhead{MHz}}
\startdata 
4 &73.0 -- 74.6                  &1.625&73.8\\
P &224 -- 480                    &16 &232,247,275,296,312,328,344,357,382,392,403,422,437,457,471\\
L &1008 -- 2032                  &64 &1040, 1488, 1808\\
S &1988 -- 3012,   2988 -- 4012  &128&2052, 2948\\
C &4188 -- 5212,   5988 -- 7012  &128&4764,   6564\\
X &7888 -- 8912,   10488 -- 11512&128&8592,  11064\\
Ku&13488 -- 14512, 15988 -- 17012&128&14192, 16564\\
K &18488 -- 19512, 24988 -- 26012&128&19065, 25564\\
Ka&31488 -- 32512, 36488 -- 37512&128&32064, 37064\\
Q &41488 -- 42512, 47488 -- 48512&128&42064, 48064\\
\enddata 
\tablecomments{The Jansky VLA electronics provides two parallel
  independently tuned analog channels.  At `P' band, only one of these
  was utilized.  At `L' band the two channels, each 512 MHz wide, were
  arranged to be be adjacent.  At all other bands, the two channels
  were separated, with frequency ranges as given in the second
  column. The `SPW Width' column gives the effective bandwidth of the
  data used for the flux density determinations.}
\end{deluxetable}

Although the new data were taken with the wideband system, not all
spectral windows were utilized, as the smoothness of the synchrotron
emission process requires only a reasonable sampling of the spectrum
to enable accurate description.  The center frequencies of the
spectral windows utilized in the analysis are given in
Table~\ref{tab:Frequencies}.  Note that all but one of the SPWs in the
P-band system were utilized\footnote{The only one not utilized, at 264
  MHz, was lost to RFI}, primarily to assist in the verification of
the new low-band system. The center frequencies given at P-band are
not uniformly separated by 16 MHz (the SPW spacing) -- this reflects
the flagging necessary to remove RFI-affected channels.

To provide an extra level of calibration, periodic observations of
three far northern sources: J0217+7349, J1153+8058, and J1800+7828,
were added at all frequency bands except `4' and `P'.  These were
included since their near-constant elevation, and 24-hour
availability, allow easier separation of temporal from elevation gain
dependencies, should the on-board gain calibration system not function
as needed.  As described below, this was indeed the case for some
bands.

\section{Observation and Calibration Methodology}

The methodology employed is identical to that described in PB13, and
will not be described in detail here.  In summary, observations of the
sources were taken in all nine VLA observing bands over the course of
a day.  Each observation was short -- less than 30 seconds -- so that
typically 6 to 10 observations (roughly, hourly) of each source, at
each band, are available for analysis.  The repetition permits
correction for various effects, such as the sensitivity and elevation
gain dependency of each band, as well as providing statistically valid
determinations of the source flux densities.  Correlator dump times
were kept short -- 1 second -- to permit detailed editing.  Editing
and calibration methods were identical to those described in PB13.

Accurate flux density measurements require accurate correction for
instrumental gain variations.  There are two principal sources of
time-variable gains for VLA observations -- temperature-induced
receiver gain changes, and elevation-dependent gains changes caused by
deformation of the primary reflector.  The first affects all bands,
the second is only appreciable at wavelengths shorter than 6cm.  At
the VLA's higher frequencies, these effects are of approximately equal
magnitude, so accurate calibration requires an electronics gain
monitor which is independent of the antenna elevation. For this
purpose, the VLA's receivers include a switched power monitoring
system whose output accurately tracks receiver gain changes.
Application of these data removes the gain variations due to the
electronics with an accuracy of $\sim$1\%, and permits accurate
measurement of the variation of antenna gain with elevation.

However, application of this gain monitioring data could not be done
for those observations which included the four strongest `absolute'
sources -- notably Cygnus A, the primary flux density calibrator -- as
it has been found that for sources such as this, which typically
multiply the total power in the system by a factor of $\sim$4, the
switched power monitor system falsely reports a small, but
significant, drop in gain.  This effect is seen in all receivers
except the low-frequency system.  Hence, in the calibration, the
switched power system was utilized only for the four highest-frequency
bands, for which no attempt to link the standard calibrators to Cygnus
A was made, and at P-band, where the false compression is not
detected.

This phenomenon was known to exist in advance of these observations,
so to allow an alternate route for removal of the amplifier gain
variations, observations of three far northern sources, whose average
elevation remains nearly constant, once every 4 hours was included.
This is a sufficient cadence, since receiver gain changes (including
the temperature-driven contribution) typically vary on timescales
longer than this.  The three sources were always observed sequentally.

Following correction for the temporal amplifier gain changes, the
elevation gain dependency, and flux densities of the target sources
was determined in the manner described in PB13.

 \subsection{Generation of Flux Densities}

 Flux densities of all source were determined through integration over
 the source brightnesses determined by standard imaging/deconvolution
 algorithms.  As shown in PB13, this straightforward method provides
 flux density values equal in reliability to those provided by other
 methods.  Atmospheric and instrumental phase fluctuations were
 corrected using phase self-calibration.  Amplitude self-calibration
 was used to identify and to remove gain fluctuations in excess of
 $\sim$5\% \footnote{To prevent bias, the corrections were not applied
   to the data.}.  At the higher frequencies, nearly all of these are
 caused by small pointing offsets, while at the low frequencies, these
 are due to instrumental gain variations which would likely have been
 corrected had we been able to utilize the switched power.  Typically,
 10 -- 20\% of the data were removed by this process, with the
 fraction rising with frequency.

 For many of the extended sources, correction of the image for
 attenuation by the primary beam is necessary.  This was done for all
 cases where the correction is more than 0.5\% using the
 recently-determined primary coefficients \citep{P16}.  We believe the
 correction is accurate to $\sim$1\% when the attenuation is less than
 10\%.  

\subsection{Placing the Scale on an Absolute Standard}

The process described above places the flux densities on a scale set
by that assumed for the primary calibrator sources.  For frequencies
above 2 GHz, we have utilized the PB13 values for 3C286, 3C295, and
3C196, whose values are directly linked to absolute measurements of
the planet Mars by WMAP, and which were shown by PB13 to be stable to
better than 1\% over the past 20 years.  For frequencies below 2 GHz,
we have used the Baars77 values for 3C405, whose values were
determined via absolutely-calibrated antennas, as reported in that
paper.  The choice of 2 GHz for the separation between these two
standards was driven by two considerations:
\begin{itemize}
\item The Mars-based calibration reported by PB13 has much
  higher errors below 2 GHz due to the weakness of Mars at these
  frequencies. Consequently, the PB13 values are much less
  secure below 2 GHz.
\item In the determination of the polynomial fits (described below)
  the residuals were significantly reduced for the three L-band
  frequencies when the Cygnus-A based values were utilized, compared
  to the Mars-based values.
\end{itemize}

\subsection{Choice of 3C405 as the Absolute Standard below 2GHz}

As noted, Baars77 provided `absolute' spectra for three sources,
Cassiopeia A, Taurus A, and Cygnus A, and a `semi-absolute' spectrum
for Virgo A\footnote{Baars77 explains that this means its spectrum was
  determined directly through ratios to the three `absolutely
  absolute' sources.}.  Of these four, only Cygnus A provides a useable
standard for absolute calibration by interferometers.  The reasons
are:
\begin{itemize}
\item Cassiopeia A and Taurus A are both supernova remnants, whose
  flux densities are expected to vary in time.  Baars77 shows that
  this secular decrease for Cassiopeia A was typically 1\%/year, and
  likely different at different frequencies.  Although little is known
  of the variability of Taurus A, we show below that its flux density
  is also significantly decreasing over time.  By contrast, Cygnus A
  is an extragalactic radio source, whose only sub-parsec component
  (the nucleus) capable of varying on timescales of interest comprises
  less than 0.1\% of the total flux density at the frequencies of
  interest.
\item At 2 arcminutes, Cygnus A is small enough to be usefully
  employed by interferometers.  Furthermore, the hotspots of Cygnus
  are very compact (arcseconds) and very bright, leading to easily
  measured visibilities on long baselines. By contrast, the angular
  extents of Taurus A, and Casseopeia A are 9.0 and 8.0 arcminutes,
  respectively, while their compact components provide a very small
  fraction of the total flux. Although Virgo A contains significant
  compact structure, its angular size of 16 arcminutes is much to
  large to be useful for calibration purposes for most
  interferometers.
\end{itemize}

\section{Polynomial Expressions for the Flux Densities}

The frequency dependencies of the spectral flux densities for all
sources were fitted (via SVD) with a polynomial function of the form
\begin{equation}
\log(S)=a_0 + a_1\log(\nu_G) + a_2[\log(\nu_G)]^2 + a_3[\log(\nu_G)]^3 + \cdots
\end{equation}
where $S$ is the spectral flux density in Jy, and $\nu_G$ is the
frequency in GHz. The choice of how many terms to use was made on the
post-fit residuals, weighted by a strong preference to use the fewest
number of terms providing a good fit. With this formalism, $a_0$ is
the log of the flux density at 1 GHz, and $a_1$ is the spectral slope
at 1 GHz.      

The resulting coefficients for all of the sources are given in
Table~\ref{tab:FitFlux}, along with the estimated errors of these
coefficients.  The fit for Fornax A is not considered reliable, as
that source is too extended to permit reliable deconvolution with the
limited data taken in this program.  Note also that 3C48, 3C138,
3C147, and 3C380 are all known or suspected variable sources, with
timescales for significant variations of a few years.  Additionally,
Taurus A and Cassiopeia A are SNRs with slowly decreasing flux
densities.  Their characteristics are discussed later in this paper.

\begin{deluxetable}{lLLLLLLLL}
\tablecaption{Fitted Coefficients for the Twenty Sources
\label{tab:FitFlux}}
\tablewidth{0pt}
\tablehead{
\colhead{Source}&\colhead{$a_0$}&\colhead{$a_1$}&\colhead{$a_2$}&
\colhead{$a_3$}&\colhead{$a_4$}&\colhead{$a_5$}&\colhead{$\chi^2$}&
\colhead{Freq. Range\tablenotemark{a}}}
\startdata
J0133-3629&1.0440\pm0.0010&-0.662\pm0.002  &-0.225\pm0.006  &       &      &      &267&0.2 - 4\\
3C48      &1.3253\pm0.0005&-0.7553\pm0.0009&-0.1914\pm0.0011&  0.0498\pm0.0009&      &      &3.1&0.05 -50\\
Fornax A  &2.218\pm0.003  &-0.661 \pm0.006 &                &&&          &17&0.2 -0.5\\
3C123     &1.8017\pm0.0007&-0.7884\pm0.0012&-0.1035\pm0.0023&-0.0248\pm0.0013&0.0090\pm0.0013&      &1.9&0.05 -50\\
J0444-2809&0.9710\pm0.0011&-0.894\pm0.004  &-0.118\pm0.010  &       &      &      &3.3&0.2 -2.0\\
3C138     &1.0088\pm0.0009&-0.4981\pm0.0022&-0.155\pm0.003  &-0.010\pm0.07&0.022\pm0.003&      &1.5&0.2 -50\\
Pictor A  &1.9380\pm0.0010&-0.7470\pm0.0013&-0.074\pm0.005  &       &      &      &8.1&0.2 -4.0\\
Taurus A  &2.9516\pm0.0010&-0.217\pm0.003  &-0.047\pm0.005  &-0.067\pm0.013&      &      &1.9&0.05 -4.0\\
3C147     &1.4516\pm0.0010&-0.6961\pm0.0017&-0.201\pm0.005  & 0.064\pm0.004&-0.046\pm0.004& 0.029\pm0.003&2.2&0.05 -50\\
3C196     &1.2872\pm0.0007&-0.8530\pm0.0012&-0.153\pm0.002  &-0.0200\pm0.0013& 0.0201\pm0.0013&      &1.6&0.050 -50\\
Hydra A   &1.7795\pm0.0009&-0.9176\pm0.0012&-0.084\pm0.004  &-0.0139\pm0.0014&0.030\pm0.003&      &3.5&0.050 -12\\
Virgo A   &2.4466\pm0.0007&-0.8116\pm0.0020&-0.048\pm0.003  &      &      &      &2.0&0.05 -3\\
3C286     &1.2481\pm0.0005&-0.4507\pm0.0009&-0.1798\pm0.0011& 0.0357\pm0.0009&      &      &1.9&0.05 -50\\
3C295     &1.4701\pm0.0007&-0.7658\pm0.0012&-0.2780\pm0.0023&-0.0347\pm0.0013& 0.0399\pm0.0013&      &1.6&0.05 -50\\
Hercules A&1.8298\pm0.0007&-1.0247\pm0.0009&-0.0951\pm0.0020&      &      &      &2.3&0.2 -12\\
3C353     &1.8627\pm0.0010&-0.6938\pm0.0014&-0.100\pm0.005  &-0.032\pm0.0005&      &      &2.2&0.2 -4\\
3C380     &1.2320\pm0.0016&-0.791\pm0.004  &0.095\pm0.022   & 0.098\pm0.022&-0.18\pm0.06&-0.16\pm0.05&2.9&0.05 -50\\
Cygnus A  &3.3498\pm0.0010&-1.0022\pm0.0014&-0.225\pm0.006  & 0.023\pm0.002&0.043\pm0.005&      &1.9&0.05 -12\\
3C444     &1.1064\pm0.0009&-1.005\pm0.002  &-0.075\pm0.004  &-0.077\pm0.005&      &      &5.7&0.2 -12\\
Cassiopeia A&3.3584\pm0.0010&-0.7518\pm0.0014&-0.035\pm0.005&-0.071\pm0.005&     &      &2.1&0.2 -4\\
\enddata
\tablenotetext{a}{The frequency range over which the coefficients are valid.}
\end{deluxetable}

\section{Estimated Errors}

\subsection{Errors in the Measured Data Values}

The errors in the determined spectral flux densities induced by the
transfer process were derived with a method different than that
utilized for PB13.  For that paper, PB13 estimated the errors for each
flux density value by deriving the variance in the $\sim$10 individual
observations whose average was used for the fit.  This procedure is
well-justified for objects which are unresolved, or only slightly
extended, as in this case each `snapshot' observation provides an
independent valid estimate for the flux density.  However, for highly
extended sources -- such as many of those included in this study --
individual snapshots do not contain enough visibility data to enable a
valid estimate of the flux density.

We have thus adopted an alternate approach which relies on the
expectation that the spectra are smooth, and that the errors are
multiplicative -- {\it i.e.}, are proportional to the flux density of
the source.  The first criterion is well justified on physical
grounds, as the emission mechanism for all these sources is
synchrotron which emits a broad, featureless spectrum extending over
many orders of magnitude \citet{Pach70}.  The second criterion is
reasonable for the sources in our list, as the SNR for all is such
that additive (thermal) noise is far less than the scatter in the
spectral fits.  Thus, the effects of gain errors (whether they be due
to receiver gain fluctuations, or pointing errors) will scale with the
source flux.  

We have thus estimated the error by computing the variance in the
ratio of the data values to the fit values for all 20 sources for each
of the 34 frequencies.  The results show that, except for three of
these 34 frequencies, the 1-$\sigma$ standard deviation is less than
0.8\%.  The three frequencies which have notably higher errors are:
232 MHz (1.6\%), 247 MHz (1.0\%), and 48 GHz (4.7\%).  All three come
from spectral windows known to be more poorly behaved than the others.
At 48 GHz, the variation is due to poor SNR and poor pointing.  At the
two lower frequencies, the high residuals are due to the poor
sensitivity at the low end of this band, combined with the flagging
needed to remove the RFI.

These variances were used in estimating the coefficients, their
errors, and the $\chi^2$ of the polynomial fits.

\subsection{Flux Density Scale Errors}

The remarkably good fits of simple polynomials to the data for all
sources strongly indicate that the effect of instrumental transfer
errors in the calibration and imaging process are much less than 1\%,
except at the highest frequency.  Thus, we believe that the error in
the proposed flux density scale is completely dominated by the errors
in the absolute standards -- Cygnus A for frequencies below 2 GHz, and
in the Mars-based flux densities above 2 GHz.

According to Baars77, the estimated error in the spectrum of Cygnus A,
between 50 MHz and 2 GHz, is $\sim$ 2 -- 4\%.  PB13 estimated the
accuracy of their scale to be 1 -- 3\%, although \citet{Par16} finds
the 42 GHz value is $\sim$ 5\% low.  We thus estimate the accuracy of
our new, comprehensive scale, to be in the range of 3 -- 5\%, with the
larger errors at the lowest and highest ends.

\begin{figure}[ht]
\centerline{\hbox{
\includegraphics[width=6.5in]{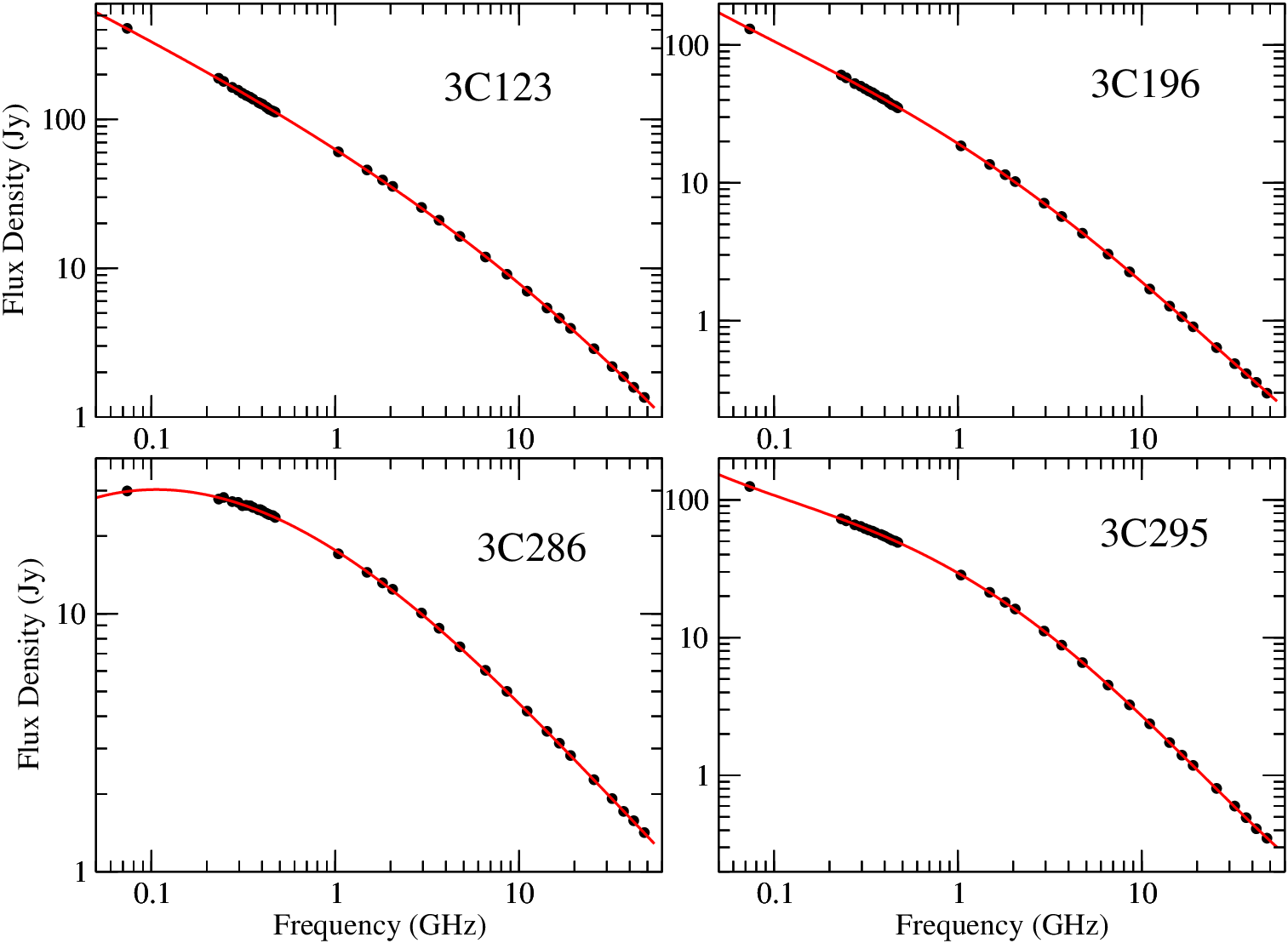}}}
\centerline{\parbox{6.5in}{
    \caption{\small Plotted are the measurements of the flux densities
      of the four identified stable standards, and the best-fit models
      for each.  For all values, the estimated errors are smaller than
      the plotted points.}
\label{fig:FourPlot}}}
\end{figure}

The new data, and the new fits, for the four sources identified as
constant to better than 1\% over 20 years are shown in
Fig.~\ref{fig:FourPlot}.

\section{Comparisons of this with other Scales}

Here we compare the values of our proposed new scale (herafter
referred to at PB16) scale with those given by the Baars77, SH12, and
PB13 scales for selected frequencies.

\subsection{Perley and Butler (PB13)}

Table \ref{tab:ScaleRatios} shows the ratio of PB16 scale values to the
PB13 scale, utilizing the four established stable sources.

\begin{deluxetable}{rcllllllll}
\tablecaption{Ratio of This Scale to the PB13 Scale 
\label{tab:ScaleRatios}}
\tablewidth{0pt} 
\tablehead{ \colhead{Source}& \colhead{328 MHz} &\colhead{1488 MHz}&
\colhead{2948 MHz}&\colhead{4764 MHz}& \colhead{8592 MHz}& \colhead{14192 MHz}&
\colhead{19064 MHz}& \colhead{32064 MHz}& \colhead{42064 MHz}} 
\startdata 
3C123&0.985&0.987&1.007&1.006&1.004&1.005&1.010&1.002&0.998\\
3C196&0.989&0.983&1.001&0.999&1.001&1.001&1.002&1.003&1.019\\
3C286&0.975&0.984&1.003&1.002&1.000&0.998&0.997&0.999&1.005\\
3C295&1.006&0.973&0.998&1.000&1.000&1.000&1.001&1.002&0.978\\
\enddata 

\tablecomments{Note that the PB13 scale is defined between 1 and 50
  GHz.  The ratios shown at 328 MHz are only to illustrate how close
  that scale comes to the new.  The 1.8\% offset noted at 1488 MHz
  reflects the transfer of the absolute standard from the Mars-based
  to the Cygnus A based absolute reference.}
\end{deluxetable}
There is excellent agreement between the PB13 and PB16 scales.  This
is expected, as the four sources are known to be time-invariant over
periods much longer than the time between these measurements.  The
average deviation of 0.982 for the 1488 MHz value reflects the change
in absolute standard, from the (poor SNR) value based on Mars of PB13,
to the Cygnus A standard adopted here.

\subsection{Scaife and Heald (SH12)}

Table \ref{tab:SHRatios} shows the ratios of the PB16 scale values to
the SH12 scale values.  

\begin{deluxetable}{rcc}
\tablecaption{Ratio of This Scale to the SH12 Scale 
\label{tab:SHRatios}}
\tablewidth{0pt} 
\tablehead{ \colhead{Source}& \colhead{73.8 MHz} &\colhead{328 MHz}}
\startdata 
3C48 &0.948&1.005\\
3C147&1.095&1.020\\
3C196&0.981&1.020\\
3C286&0.955&1.091\\
3C295&1.005&1.037\\
3C380&0.950&0.983\\
\enddata 
\tablecomments{The SH12 scale is defined between 30 and 300 MHz.}
\end{deluxetable}
The SH12 scale is shown to be quite close to the absolute, with only
3C286 showing more than $\sim$5\% discrepancy.  This might be due to
the presence of unusually close confusing sources.  The low value in
the flux density for 3C380 can be attributed to its known variability,
see the discussion below.

\subsection{Baars et al. (Baars77)}

Table \ref{tab:BaarsRatios} shows the ratio of the PB16 scale to that
of Baars77.

\begin{deluxetable}{rclllll}
\tablecaption{Ratio of This Scale to the Baars77 Scale 
\label{tab:BaarsRatios}}
\tablewidth{0pt} 
\tablehead{ \colhead{Source}& \colhead{328 MHz} &\colhead{1488 MHz}&
\colhead{2948 MHz}&\colhead{4764 MHz}& \colhead{8592 MHz}& \colhead{14192 MHz}} 
\startdata 
3C48 &0.978&1.025&1.017&1.006&1.009&1.036\\
3C123&1.074&0.991&0.973&0.948&0.925&0.910\\
3C144&0.777&0.888&0.904&     &     &\\
3C147&1.010&0.993&0.967&0.951&0.964&1.022\\
3C218&0.999&1.027&0.970&0.933&$0.886^*$&\\
3C274&0.907&1.000&     &     &     &\\
3C286&0.977&1.013&1.004&0.987&0.976&0.979\\
3C295&1.001&1.012&0.999&0.980&0.974&0.997\\
3C348&0.969&1.057&1.038&0.994&$0.925^*$&\\
3C353&1.034&1.007&1.038&     &     &\\
3C405&1.000&1.006&0.975&0.977&1.009&\\
3C461&0.820&0.846&0.821&     &     &\\
\enddata 
\tablecomments{Shown are ratios where both the Baars scale is defined
  and for which we have data.  Values marked with an asterisk are
  known to be too low, due to overresolution by the VLA. }
\end{deluxetable}
There is in general good agreement between the scales for those
frequencies where the more extended sources are not heavily resolved
by the VLA observations.  The much lower values for 3C144 and 3C461
reflect the decline in the flux density of these two supernova
remnants over time.  The variations for 3C48 and 3C147 can be
attributed to these sources' known slow variability.  The drop in flux
density for 3C123 is more surprising -- this is discussed below.

\section{Comments on the Sources}

Here we present short summaries of each of the target sources, with
emphasis on the suitability of each for radio telescope calibration
purposes.  
\subsection{3C123}
This source, a radio galaxy at redshift $z=0.218$, is one of the
unchanging sources identified by PB13.  Its angular size of 44
arcseconds makes it of limited use for high resolution
interferometers.  See PB13 for an image with 3 arcsecond resolution.
Roughly speaking, it is too heavily resolved on baselines exceeding
200K$\lambda$ to be used as a calibrator.  In VLA terms, this
corresponds to frequencies above 2 GHz, although it could be used in
the compact C and D configurations up to 12 GHz.

The new data show no evidence of any secular change (to $\sim 1\%$)
in flux density since 2012.  However, comparison (from Table 8) to the PB13
scale shows that current flux density is now considerably lower than
that given in Baars, from 1\% at 1488 MHz to 9\% at 14.2 GHz.  It is
very unlikely that the Baars scale could be this much in error, and we
suggest that the nucleus of the source is now much less strong than it
was prior to 1975.  If real, the decline is significant, as the
nuclear emission flux density is currently about 100 mJy from 1.5
through 15 GHz, a drop of about 400 mJy from the levels needed to
reconcile these measurements to the scale of Baars77.  However, this
hypothesis is unlikely to explain the 7\% rise in the flux density at
328 MHz -- although it should be kept in mind that the Baars77 scale
is not valid at that frequency.  
\subsection{3C196}
This source, a quasar at redshift $z = 0.871$ is a $\sim$ 7 arcsecond
double with compact lobes. The structure is shown in PB13.  Table 6
shows no change in flux density to an accuracy well better than
1\%. The nuclear core is extremely weak -- less than $\sim$2 mJy at
all frequencies.  The lobe structures, although compact, are of
kiloparsec scale, ensuring no significant change in flux density on
long timescales.  The small size, high flux density and simple
spectrum make this source an excellent calibrator at low frequencies.  
\subsection{3C286}
This extraordinary source, a quasar at redshift $z=0.846$ is very
compact, with weak, steep-spectrum structures extending about 3
arcseconds to the west, and 0.5 arcseconds to the east of the nuclear
core, as shown in PB13.  The absence of any detectable variabililty
in both the total flux density and polarized emission is likely linked
to the absence of an inverted spectrum core -- a most unusual --
perhaps unique -- feature for such a compact object.
\subsection{3C295}
This is a radio galaxy at redshift $z=0.464$, with a simple
double-lobed structure of angular size 5 arcseconds, and weak nuclear
core of about 5 mJy.  There is no sign of variability in the source,
as expected, as the nuclear emission at all frequencies is less than
1\% of the total.
\begin{figure}[ht]
\centerline{\hbox{
\includegraphics[width=6.5in]{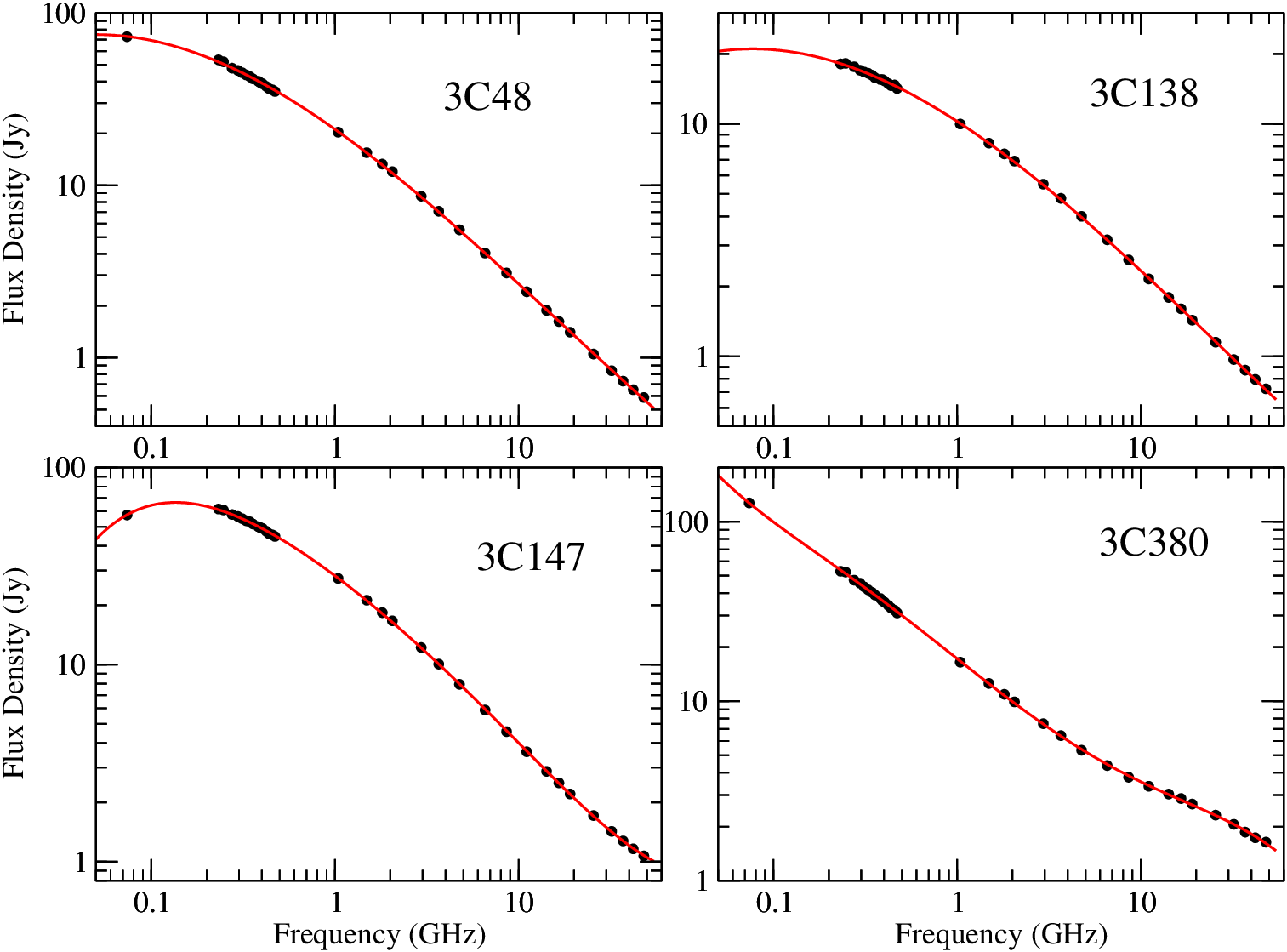}}}
\centerline{\parbox{6.5in}{
    \caption{\small Showing the spectra and fits for 3C48, 3C138,
      3C147, and 3C380.  Errors are smaller than the plotted points.
      The inflection in the spectrum for 3C380 seen at the highest
      frequencies is due to the strong, flat-spectrum nuclear
      emission.}
\label{fig:FourPlotB}}}
\end{figure}

\subsection{3C48}
This source, a compact (1.2 arcseconds maximum extent) steep-spectrum
quasar of redshift $z=0.367$ was shown to be slowly variable in PB13.
Its structure is shown in PB13.  Comprison of the current values to
those of PB13 show no change exceeding 2\% between 1 and 50 GHz.  Its
spectrum is shown in Fig~\ref{fig:FourPlotB}.
\subsection{3C138}
PB13 showed that this source -- a CSS quasar of redshift $z=0.759$
with maximum extent $\sim$ 0.7 arcseconds has undergone a significant
flare from its nuclear core, starting in 2002, peaking in 2010, and
declining therafter.  The new observations show the flux density
continuing to decline over all frequencies, with current (2016) values
some 5 -- 16\% lower than in 2012.  Because of these significant
variations, this source is not recommended as a flux density
calibrator, although its high polarization appears to be rather more
stable. Its spectrum is shown in Fig~\ref{fig:FourPlotB}, and its
structure in PB13.
\subsection{3C147}
PB13 showed that this source has had sporadic flux density changes
from its nuclear core of up to 20\% on short (few years) timescales
over the period 1980 through 2012.  The recent data show a sudden
increase in flux density at the highest frequencies -- from barely 1\%
at 2 GHz to over 20\% at 48 GHz. As there is little change in flux
density between our October 2014 and January 2016 observations, this
increase must have occured between January 2012 and October 2014.  Its
spectrum is shown in Fig~\ref{fig:FourPlotB} and its structure -- with
maximum extent of $\sim$ 0.9 arcseconds, is shown in PB13.
\subsection{3C380}
\begin{figure}[!ht]
\centerline{\hbox{
\includegraphics[width=3.25in]{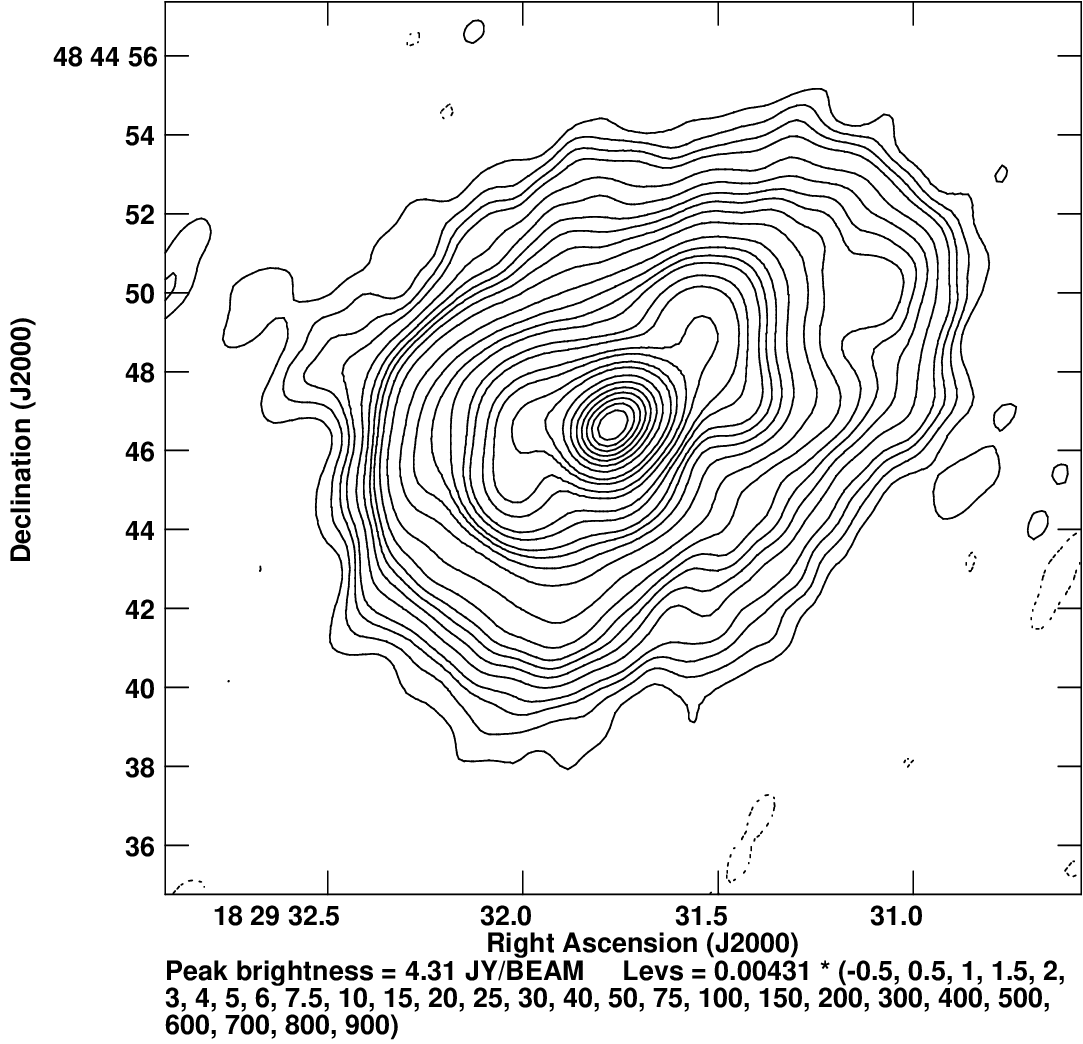}
\includegraphics[width=3.25in]{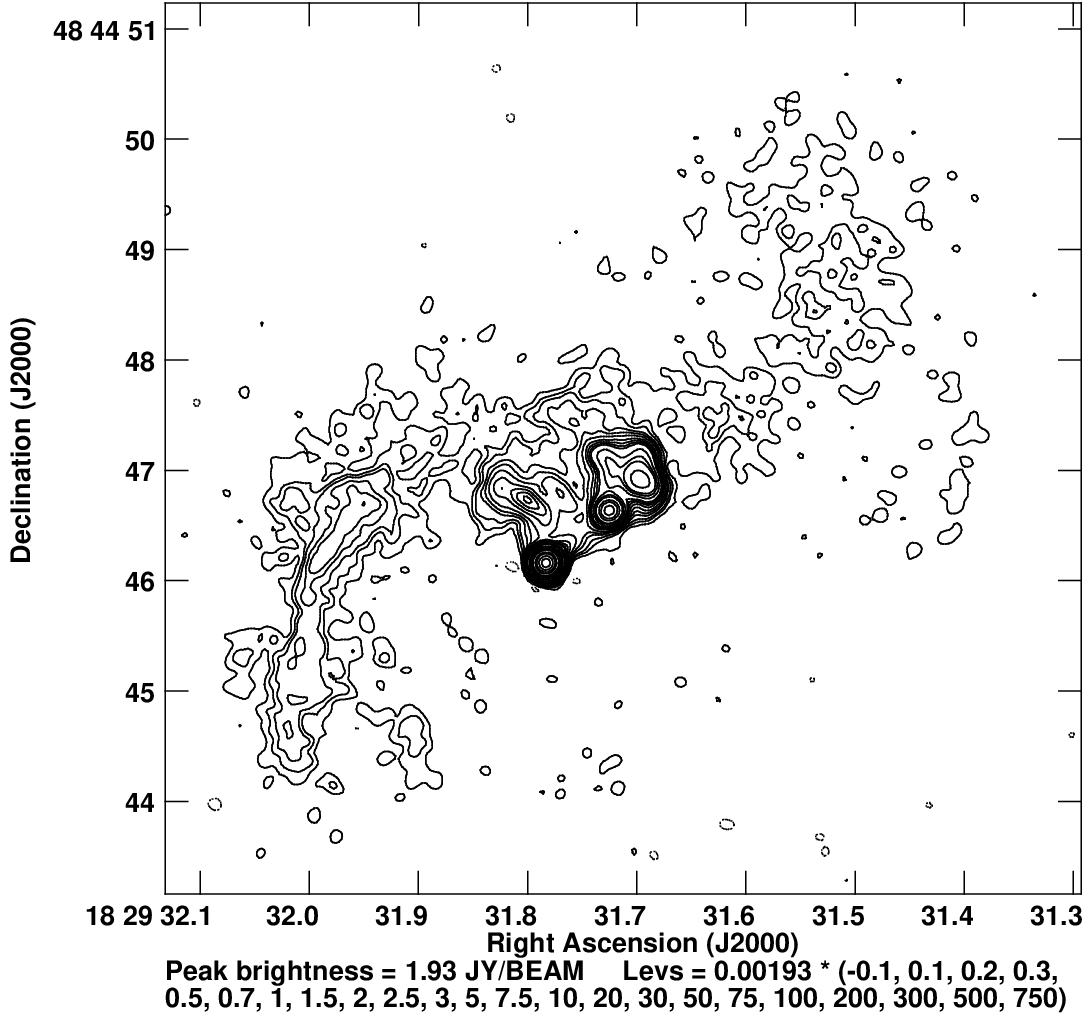}
}}
\centerline{\parbox{6.5in}{
    \caption{\small Images of 3C380.  (Left) The structure at 1488 MHz
      with 1.4 arcsecond resolution.  The smooth elliptically-shaped
      halo has dimension 16 x 12 arcseconds.  (Right) The structure at
      16564 MHz with 0.125 arcsecond resolution cannot detect the
      halo, and is dominated by a complex of very compact
      features. The more northern of the two compact features is the
      nucleus.  }
\label{fig:3C380}}}
\end{figure}

This source, a CSS (Compact Steep Spectrum) quasar at redshift
$z=0.691$ was included in the new observations since is is included in
the list of SH12.  It has complex structure, showing both a diffuse
elliptical halo, and a central complex comprising a number of very
compact sources, including a strong, unresolved nucleus, as shown in
Figure~\ref{fig:3C380}.  The central highest-brightness feature is the
flat spectrum nucleus. Its overall source spectrum (for 2016) is shown
in Fig~\ref{fig:FourPlotB}.
The emission at higher frequencies is increasingly dominated by the
flat-spectrum nucleus.  VLBA observations of this component by
\citet{List13} show a highly superluminal jet ($\beta \sim 13.1$)
emanating to the NW from the core.  As expected by the presence of a
strong core and milliarcsecond jet, this source is highly variable --
the January 2016 data show the flux density to have declined from the
October 2014 value by $\sim$3\% at 16.6 GHz to nearly 30\% at 48
GHz. Because of this, and the size and complexity of the structure,
this object is only of value for calibration at the lowest
frequencies, where the variability is much reduced, and the angular
scale of the halo not of concern.
\begin{figure}[!ht]
\centerline{\hbox{
\includegraphics[width=6.5in]{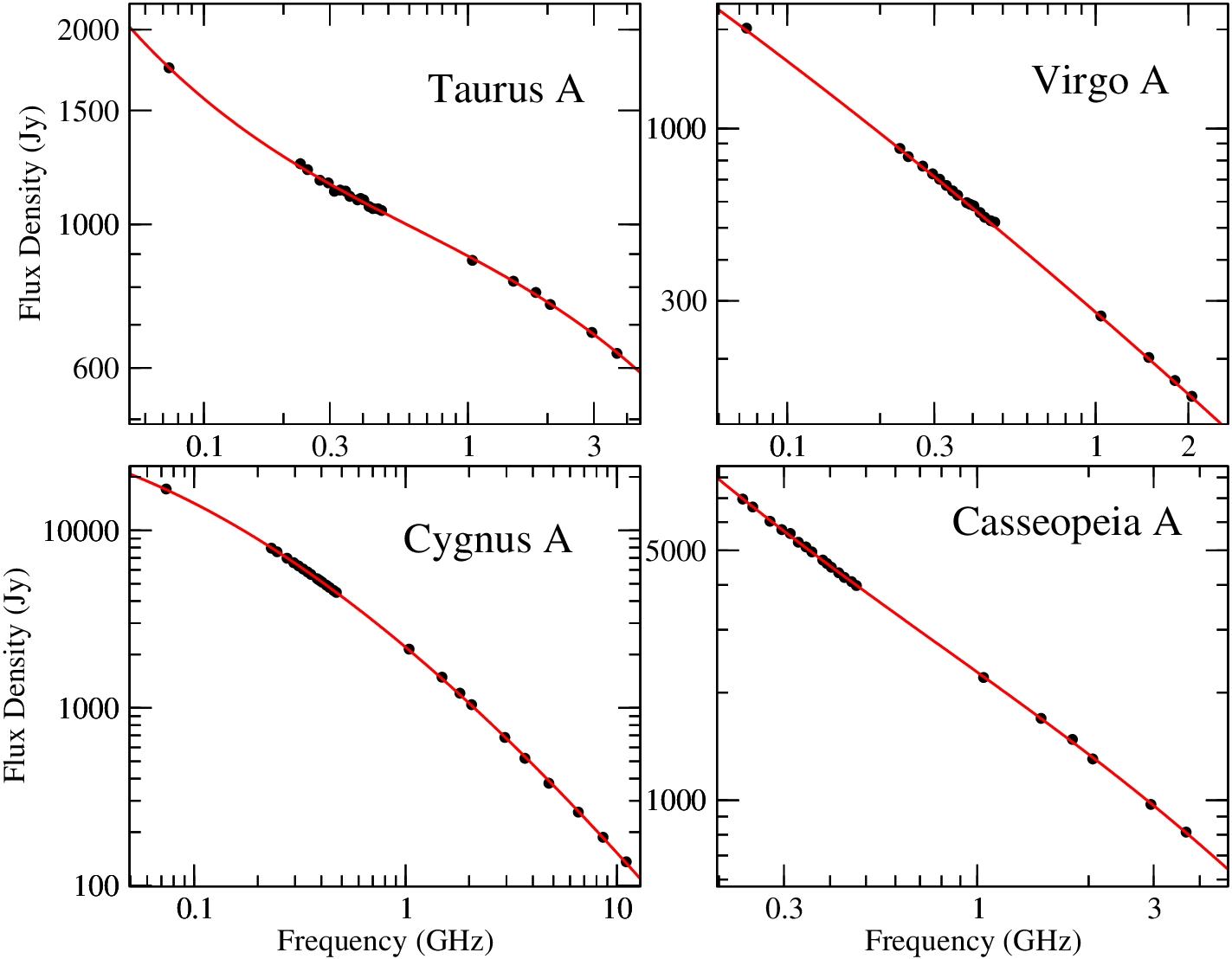}}}
\centerline{\parbox{6.5in}{
    \caption{\small The spectra and fits for the four Baars77
      `absolute' standard sources.  The plotted values for Taurus A
      and Casseopeia A -- both variable sources -- are from 2016.}
\label{fig:FourPlotC}}}
\end{figure}
\subsection{Taurus A}
\begin{figure}[!ht]
\centerline{\hbox{
\includegraphics[width=3.25in]{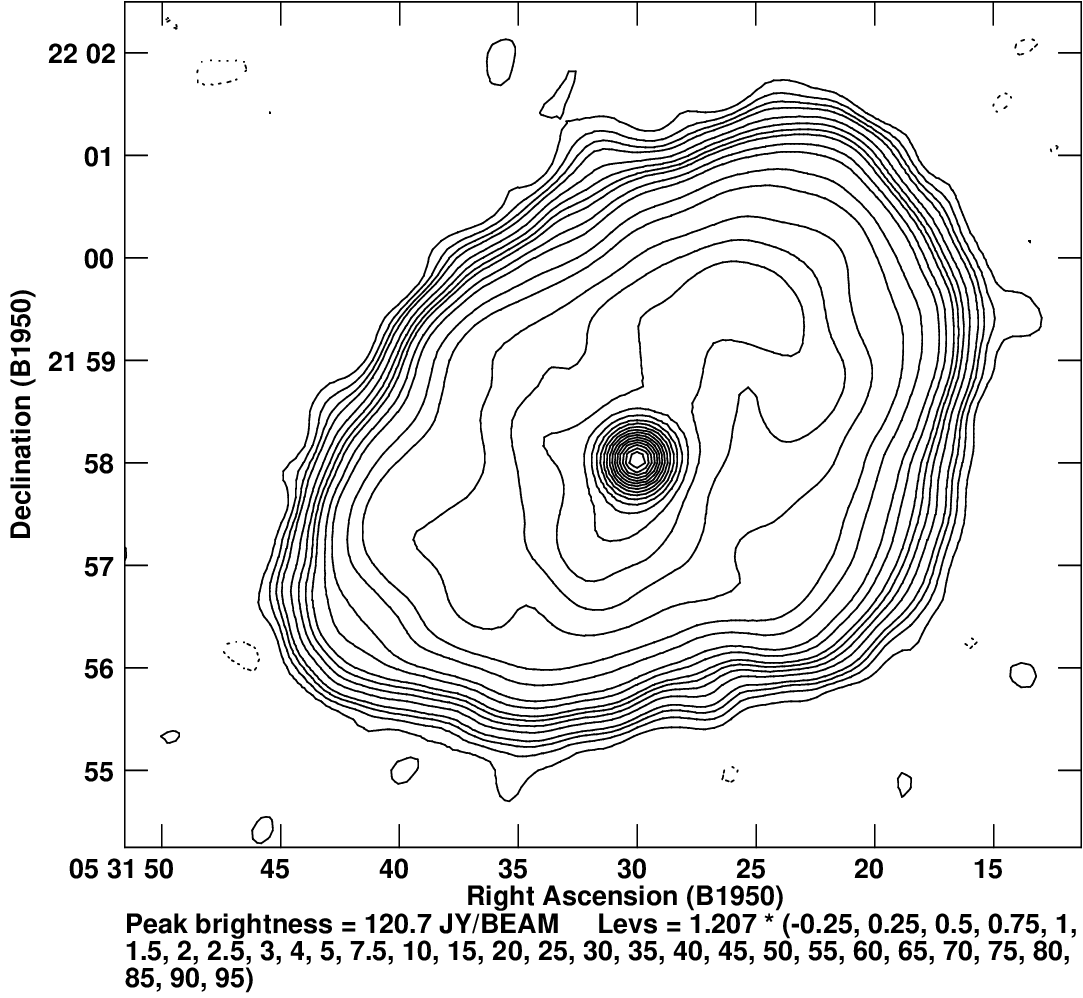}
\includegraphics[width=3.25in]{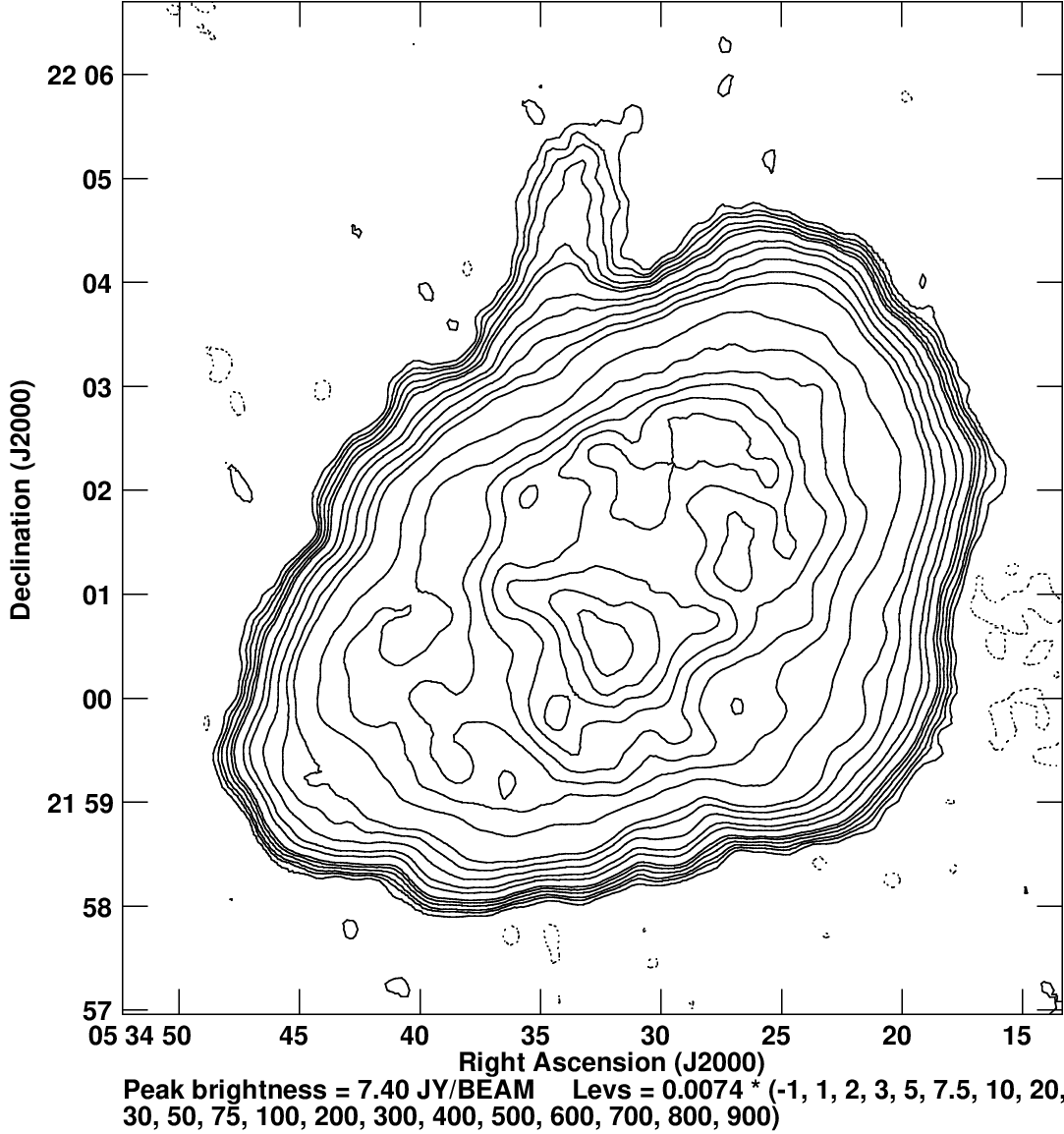}
}}
\centerline{\parbox{6.5in}{
    \caption{\small Taurus A (Crab Nebula, 3C144) at 73.8 MHz with 30
      arcseconds resolution (left), and at 1488 MHz with 17 arcseconds
      resolution (right).  The major difference is the pulsar, whose
      averaged 73.8 MHz emission of 90 Jy is easily visible in the
      left panel, but is not perceptible at 1488 MHz.  The plume,
      easily visible at 1040 MHz, is at the noise level in the 73.8
      MHz image.}
\label{fig:Taurus}}}
\end{figure}

This famous radio source is a SNR and pulsar wind nebula resulting
from a bright supernova in the year 1054 AD.  The central pulsar is
powering the synchrotron emission, and the radio-emitting nebula is
expanding outwards at about 0.15 arcsecond/year \citep{B91}.  Low
resolution images of the source at 1488 MHz and 73.8 MHz are shown in
Fig.~\ref{fig:Taurus}. Its spectrum is shown in
Fig~\ref{fig:FourPlotC}.
The observed expansion, and presence of the pulsar strongly suggest
that the source may be variable.  We can find little information on
any measured secular change of the flux density.  However, the current
flux density between 1 and 4 GHz is 10 -- 15\% lower than the values
given by Baars77.  As it is very improbable that the latter value
be in error by this much, the indication is that the source's flux
density is declining by approximately 0.25\%/year.

The large extent, lack of small angular scale structure, and secular
changes all argue that this source is unsuitable as a flux density
standard.
\subsection{Virgo A}
\begin{figure}[!ht]
\centerline{\hbox{
\includegraphics[width=3.25in]{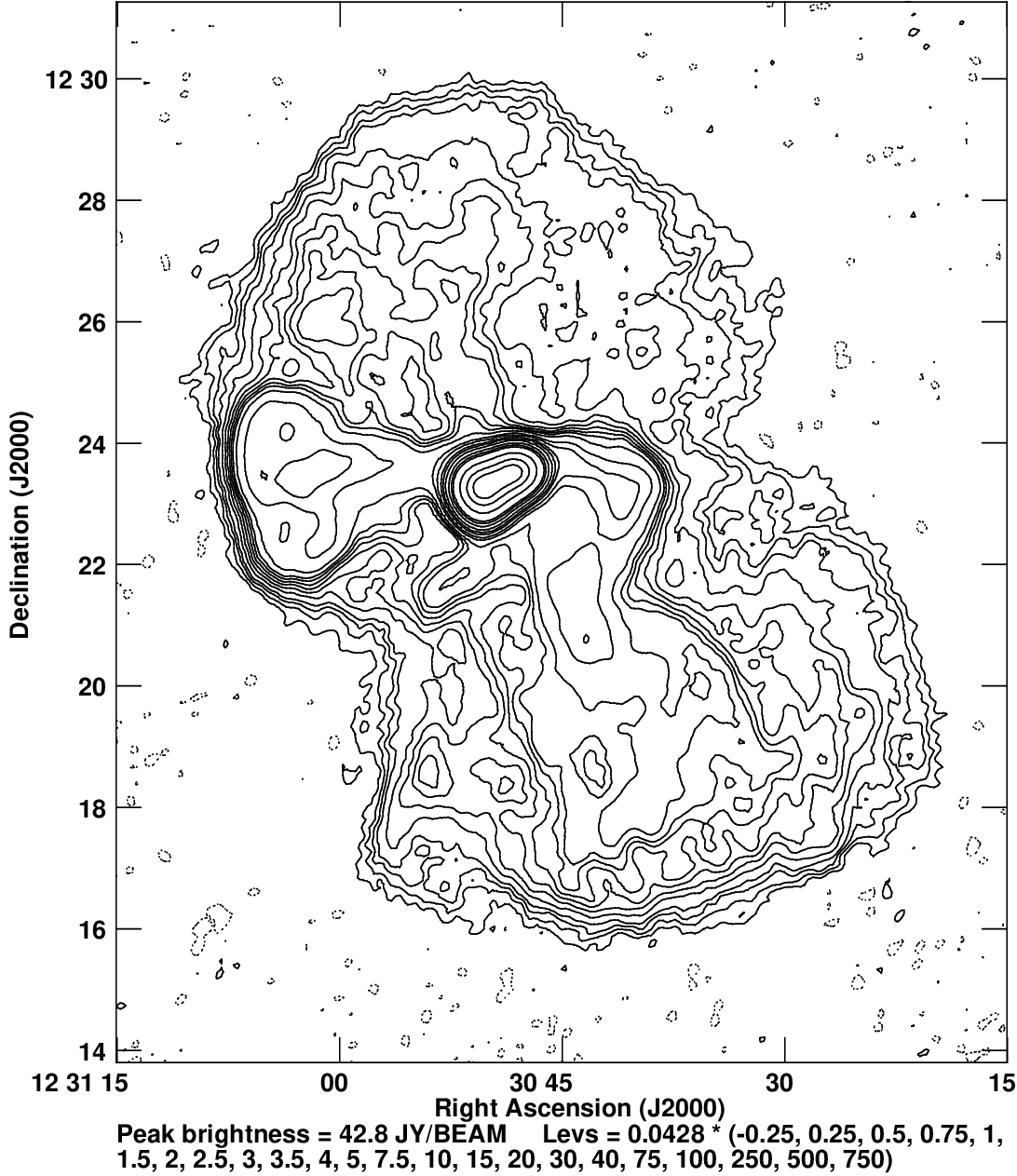}
\includegraphics[width=3.25in]{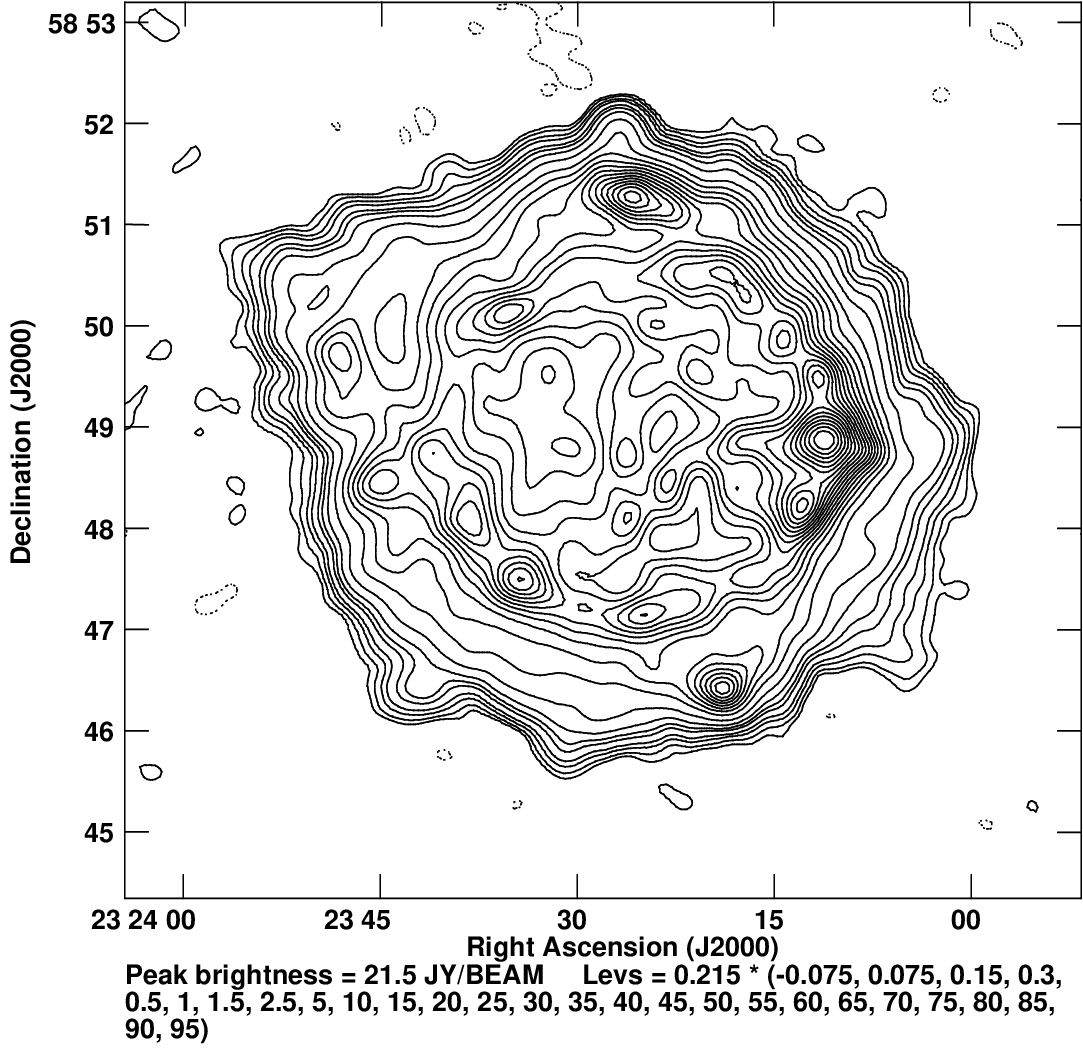}
}}
\centerline{\parbox{6.5in}{
\caption{\small (Left)  Virgo A at 1488 MHz, with 24 arcseconds
  resolution.  The compact region in the center contains the
  well-studied one-sided jet.  The diffuse lobes extend over 14
  arcminutes.  (Right) Cassioepeia A at 1488 MHz, with 17 arcseconds
  resolution. }
\label{fig:VirgoCass}}}
\end{figure}

The radio emission from this well-known radio galaxy is very well
studied thanks to its proximity (16 Mpc) and angular size (14
arcminutes).  Besides the one-sided jet, the source is comprised of
very faint and extended radio lobes, as shown in
Fig.~\ref{fig:VirgoCass}.  It is the presence of these very large and
diffuse structures that make this source very problematical for
calibration purposes, except for low resolution arrays and
single-dishes.  The current flux density from 1 to 4 GHz is within 1\%
of the Baars77 value.  However, at P-band, the newly measured values
are low by about 5\% compared to Baars77.  Its spectrum is shown in
Fig~\ref{fig:FourPlotC}.
\subsection{Cassiopeia A}
This very strong radio source is identified with an unseen SN
approximately 300 years old. It has long been identified as the
strongest extra-solar radio source.  However its continued secular
decline has relegated it two 2nd place, behind Cygnus A, for
frequencies below $\sim$ 1 GHz.  The structure at 1488 MHz, with 17
arcseconds resolution, is shown in Figure~\ref{fig:VirgoCass}. Its
spectrum (2016) is shown in Fig~\ref{fig:FourPlotC}. Its large angular
size and generally diffuse structure make this a poor object for
calibration purposes.  The flux density is declining at a rate given
by Baars77 as $-0.97 + 0.3\log(\nu_G)$ percent/year.  However, our
recent measurements indicate a much slower decline: the 74 MHz flux
density measured in 1998 is down by 9\% from the 1977 value -- a
0.41\%/year decline.  The 2016 flux density, from 230 through 4000 MHz
is 18\% lower than the Baars77 value for 1977 -- a 0.46\%/year
decrease.
\subsection{Cygnus A}
\begin{figure}[!ht]
\centerline{\hbox{
\includegraphics[width=3.25in]{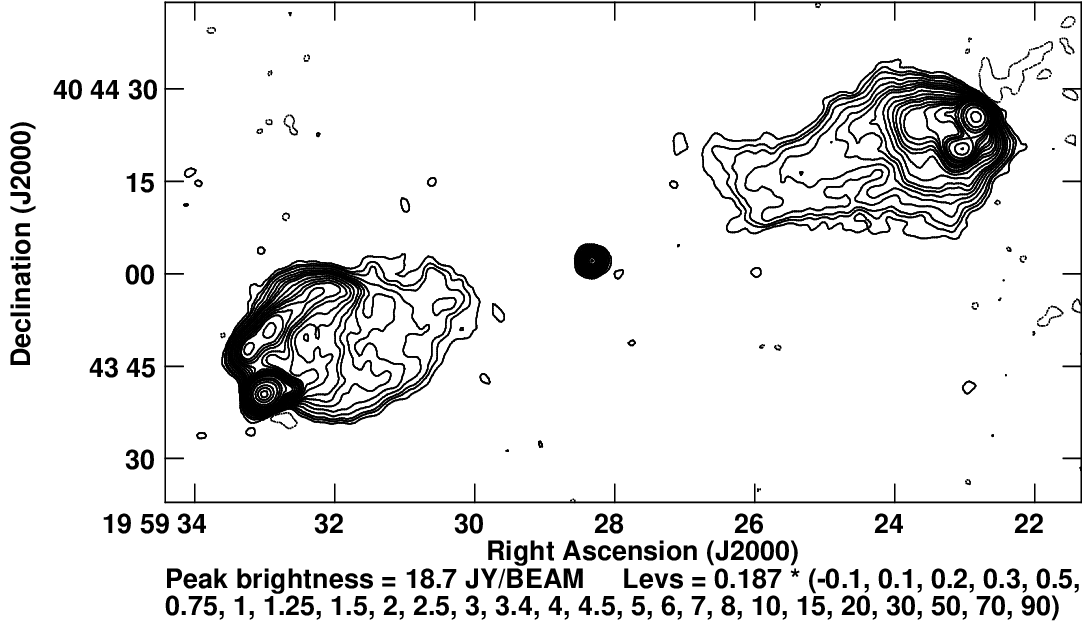}
\includegraphics[width=3.25in]{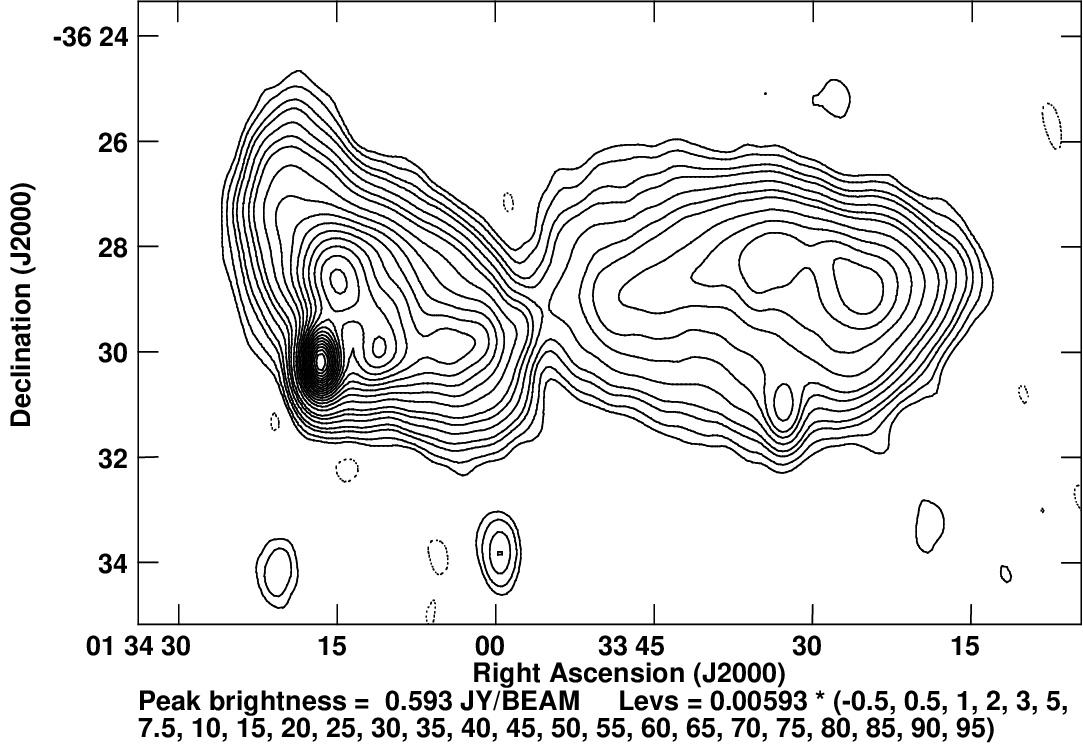}
}}
\centerline{\parbox{6.5in}{
\caption{\small (Left) Cygnus A at 11.06 GHz, with 2.25 arcseconds
  resolution.  The source has a maximum extent of 130 arcseconds, with
  bright and compact hotspots.  (Right) The southern source
  J0133-3629, at 1.04 GHz with 60 x 30 arcseconds resolution.}
\label{fig:Cygnus}}}
\end{figure}

Cygnus A (3C405) is a nearby ($z=0.056$) luminous radio galaxy.  Due
to its proximity and very high flux density, its structure has been
extensively studied.  An image at a frequency of 11 GHz with 2.25
arcsecond resolution is shown in Figure~\ref{fig:Cygnus}.  With a
maximum extent of just 2 arcminutes, and with very bright and sharp
arcsecond-scale spatial features, the source is a good calibrator for
objects at low frequencies to moderate-resolution arrays. Its spectrum
is shown in Fig~\ref{fig:FourPlotC}. Other than the 1 Jy nuclear
source, the most compact features have physical extents of hundreds of
parsecs, so that secular changes in flux density on decadal timescales
must be at a very low fractional level.

The remaining eight objects are all southern sources, and were
included in this study as an aid to transferring the northern
calibrator network to the southern hemisphere.

\begin{figure}[!ht]
\centerline{\hbox{
\includegraphics[width=6.5in]{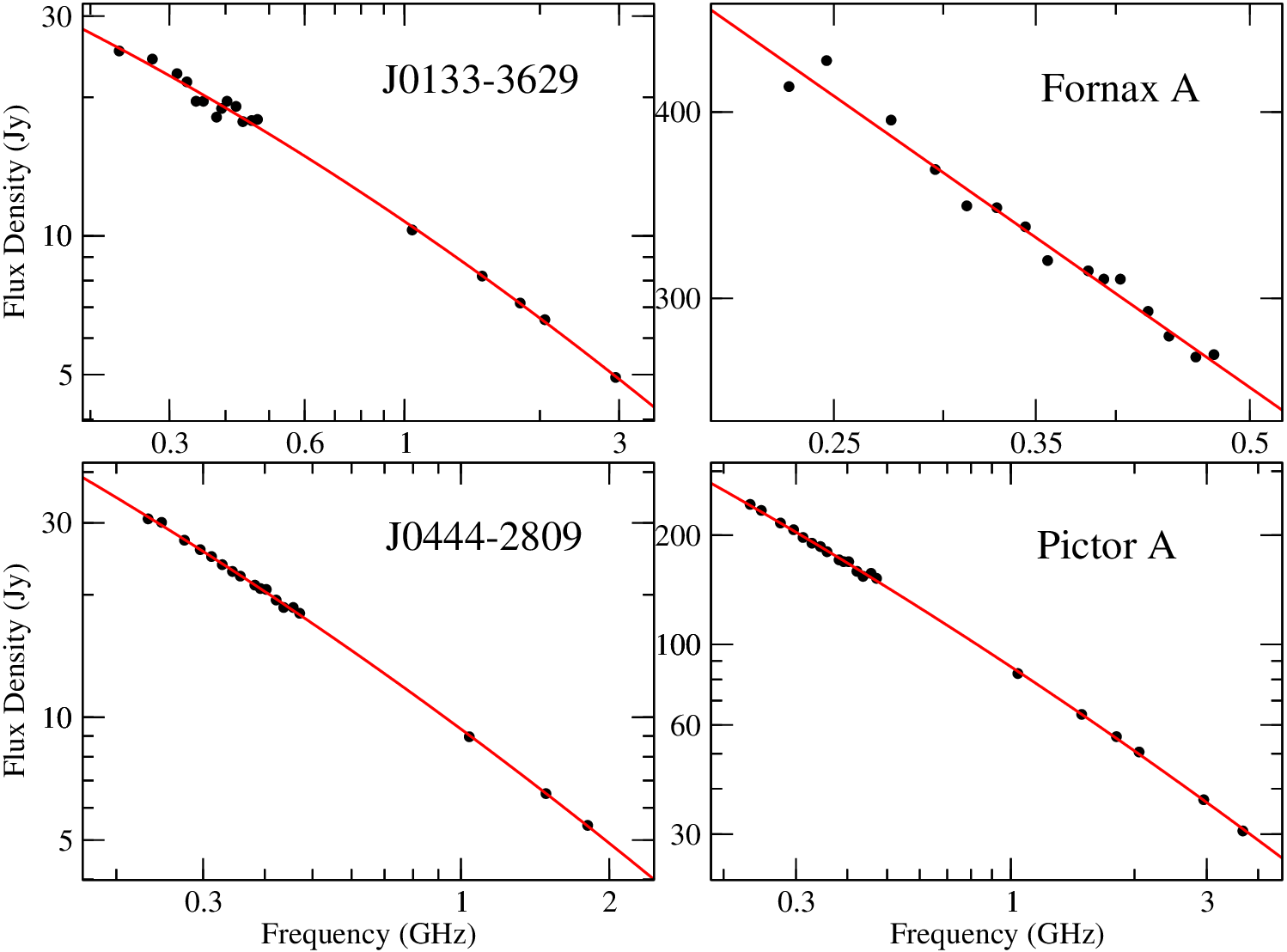}}}
\centerline{\parbox{6.5in}{
\caption{\small Spectra of four of the southern calibrators. The poor
  fits for J0133-3629 and Fornax A reflect the difficult in accuracy
  reconstruction of large, far southern sources with just a few VLA
  snapshot observations.}
\label{fig:FourPlotD}}}
\end{figure}

\subsection{J0133-3629}
This source is a very large (14 arcminutes) and diffuse double lobed
object with a bright compact hotspot in the eastern lobe.  The nuclear
emission from this source is weak (30 mJy) at 2.9 GHz), so that
measureable variability in the total flux density is unlikely.
Nevertheless, this is a poor object for use by interferometers for
flux density calibration, due to its large size and diffuse structure.
Its spectrum is shown in Fig~\ref{fig:FourPlotD}.
\begin{figure}[!ht]
\centerline{\hbox{
\includegraphics[width=3.25in]{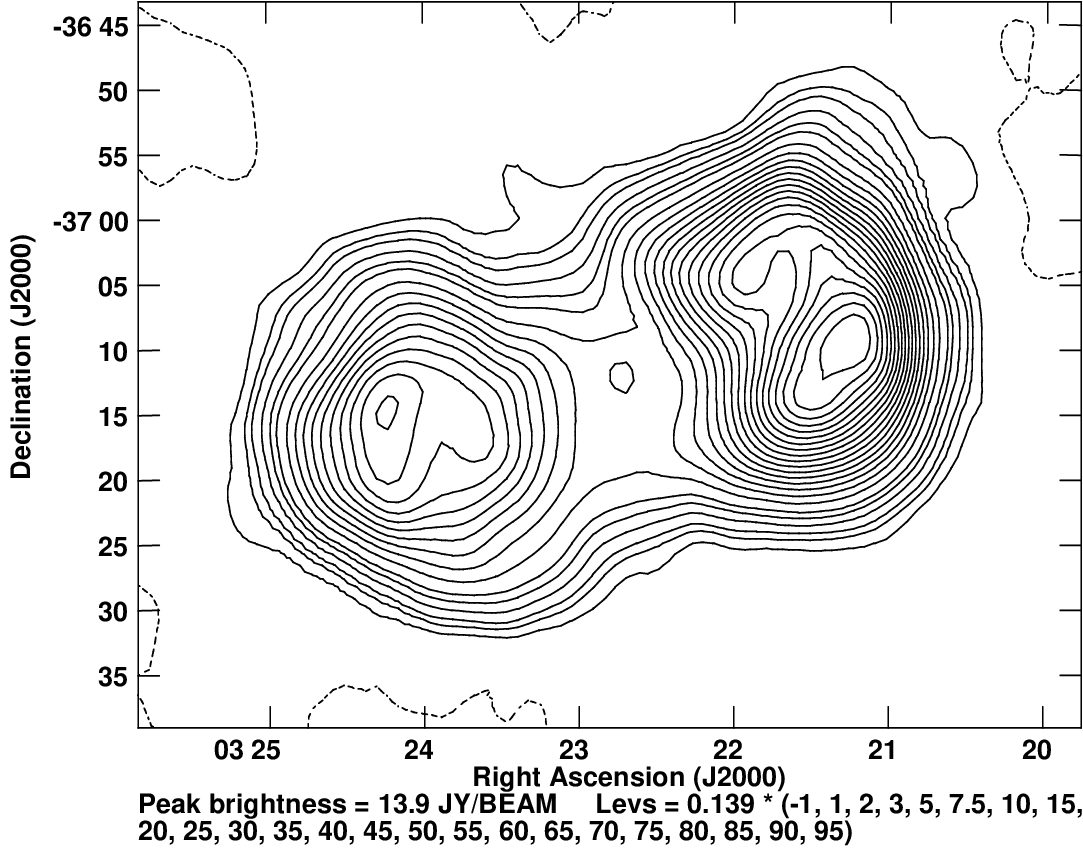}
\includegraphics[width=3.25in]{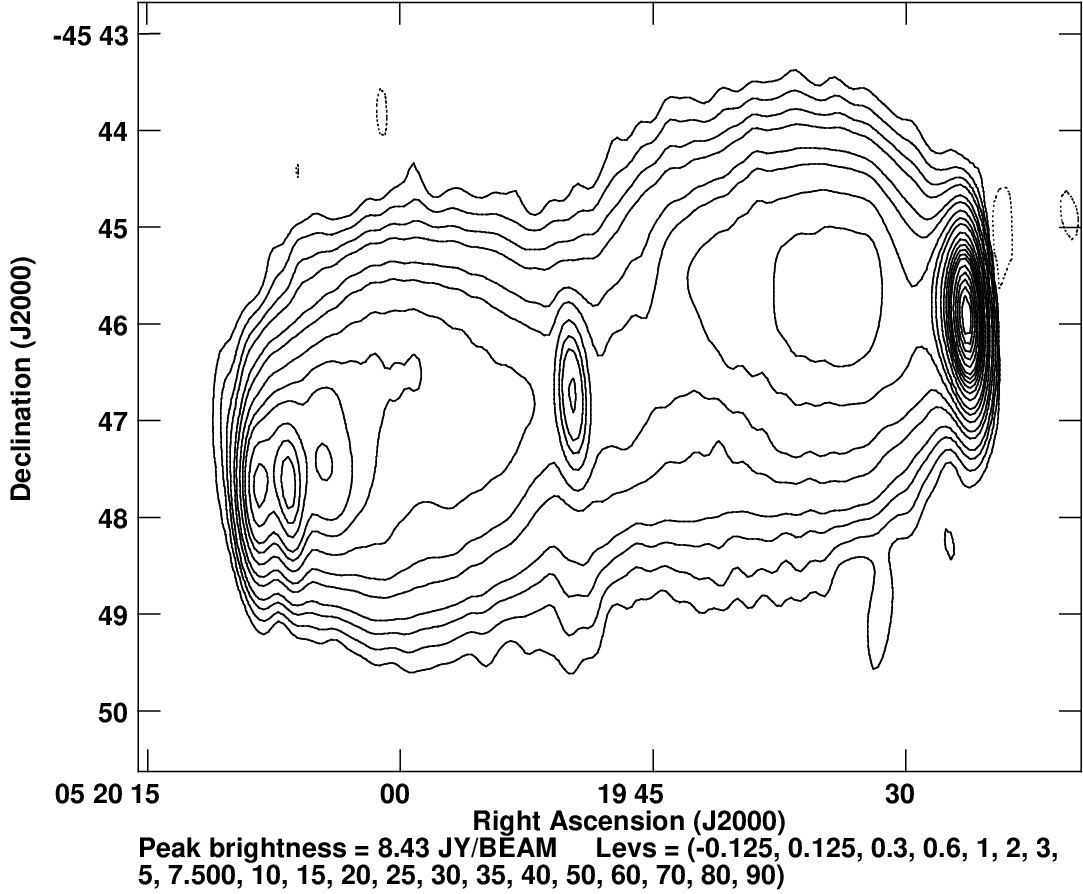}
}}
\centerline{\parbox{6.5in}{
\caption{\small (Left)  Fornax A at 312 MHz with 200 x 275 arcseconds
  resolution.  This object is too large to be properly imaged with
  only a few VLA snapshots.  (Right) Pictor A at 1808 MHz and 14 x 65
  arcseconds resolution.  }
\label{fig:FornaxPictor}}}
\end{figure}
\subsection{Fornax A}
This source, associated with the galaxy NGC1316, is a nearby (19 Mpc)
large (55 arcminutes) and extremely diffuse radio galaxy.  Its angular
size is too large to enable the VLA to make accurate flux density
measurements above 1 GHz.  The limited snapshot observations of this
program are not sufficient to permit accurate reconstruction of its
structure, so that the integrated flux densities, and the spectrum
shown in Fig~\ref{fig:FourPlotD} must be viewed with considerable
scepticism.  Figure~\ref{fig:FornaxPictor} shows a low-resolution image
at 312 MHz, with 200 x 275 arcsecond resolution.  This source is too
large and diffuse to be useful as a primary calibrator for
high-resolution interferometers. 
\subsection{Pictor A}
This FRII radio galaxy is a large (8.3 arcminutes) and strong source.
An image at 1808 MHz with 14 x 65 arcsecond resolution in shown in
Figure~\ref{fig:FornaxPictor}.  Its spectrum is shown in
Fig~\ref{fig:FourPlotD}.  The prominent hotspots make this a useful
object for interferometric calibration.  The compact nuclear source
contributes less than 0.5\% of the total flux density at P-band, so
any secular variations of this component at that frequency will have
negligible effect on the total flux density.  At higher frequencies,
its large angular size, and the increasingly prominent nucleus makes
this source unsuitable for calibration purposes.  
 
\subsection{J0444-2809}
\begin{figure}[!ht]
\centerline{\hbox{
\includegraphics[width=3.25in]{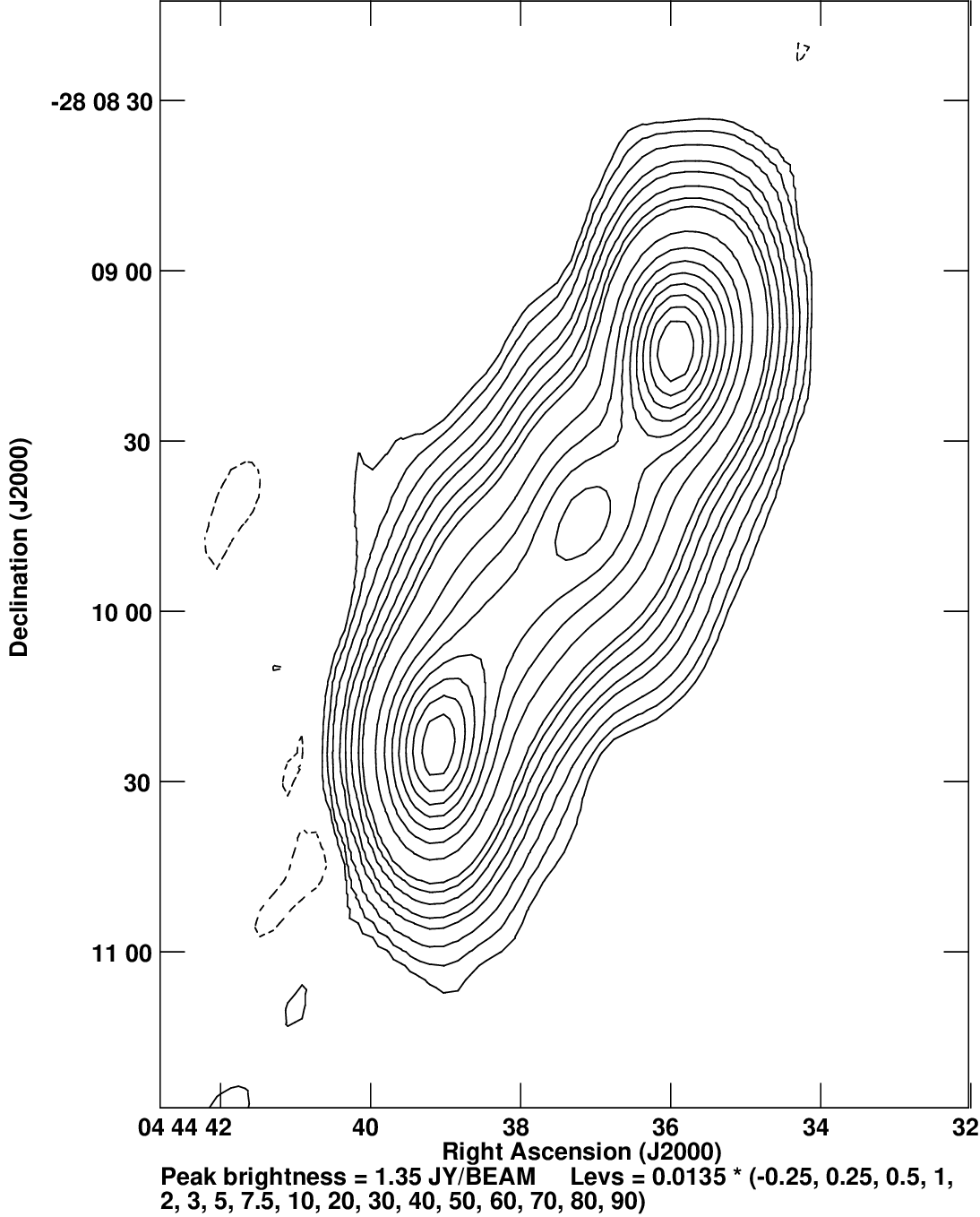}
\includegraphics[width=3.25in]{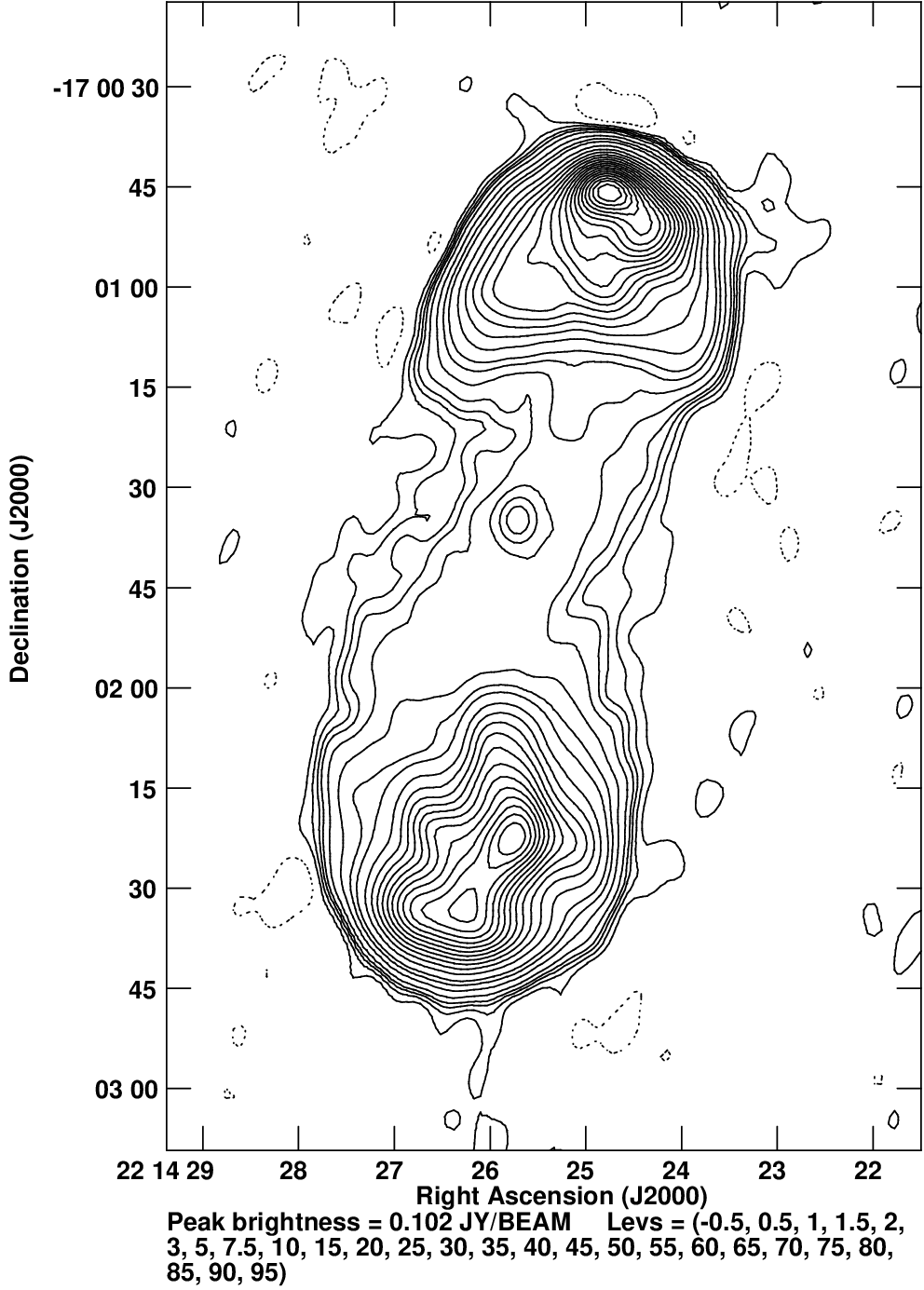}
}}
\centerline{\parbox{6.5in}{
\caption{\small Images of J0444-2809 at 1808 MHz with 25 x 13
  arcsecond resolution, and 3C444, at 4764 MHz with 7 x 5 arcsecond resolution.}
\label{fig:Both444}}}
\end{figure}

This double-lobed radio source is relatively compact (2 arcminutes)
and strong, making it a potentially useful calibrator.  A
low-resolution image (25 x 13 arcseconds) at 1808 MHz is shown in
Figure~\ref{fig:Both444}.  Its spectrum is shown in
Fig~\ref{fig:FourPlotD}. Due to a scheduling error, we did not observe
this potentially useful calibrator at a higher frequency.
\subsection{3C444}
\begin{figure}[!ht]
\centerline{\hbox{
\includegraphics[width=6.5in]{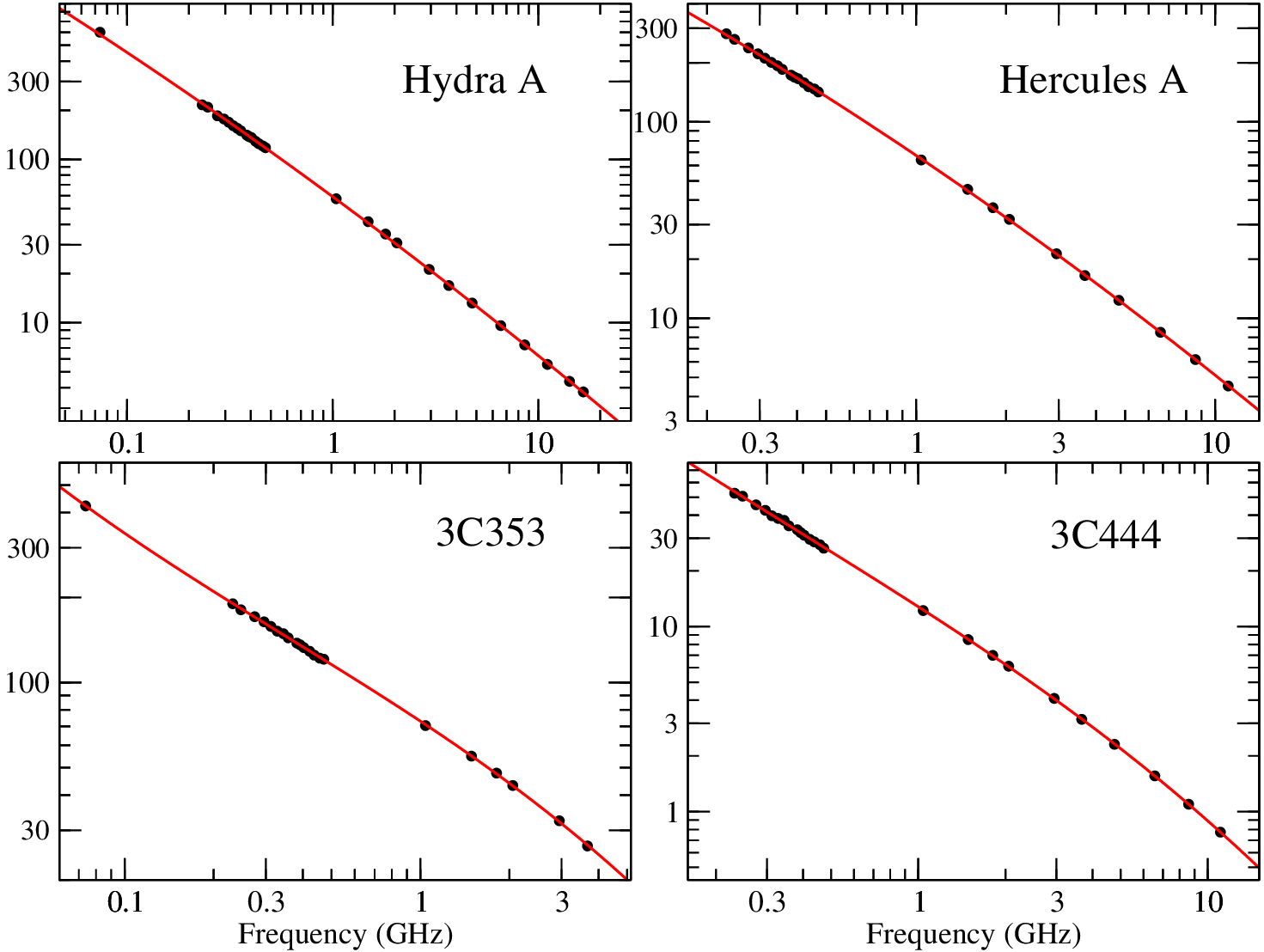}}}
\centerline{\parbox{6.5in}{
\caption{\small The observed data and LSQ fits for Hydra A, Hercules
  A, 3C353, and 3C444.  Errors are smaller than the plotted points.}
\label{fig:FourPlotE}}}
\end{figure}

An image of this $z=0.153$ radio source, at 4764 MHz with 7 x 5
arcsecond resolution is shown in Figure~\ref{fig:Both444}.  Its
spectrum is shown in Fig~\ref{fig:FourPlotE}.  Its maximum size of two
arcminutes, and absence of a prominent core, make it a useful
calibrator for low-frequency interferometers of moderate resolution.  

\subsection{Hydra A}
\begin{figure}[!ht]
\centerline{\hbox{
\includegraphics[width=3.25in]{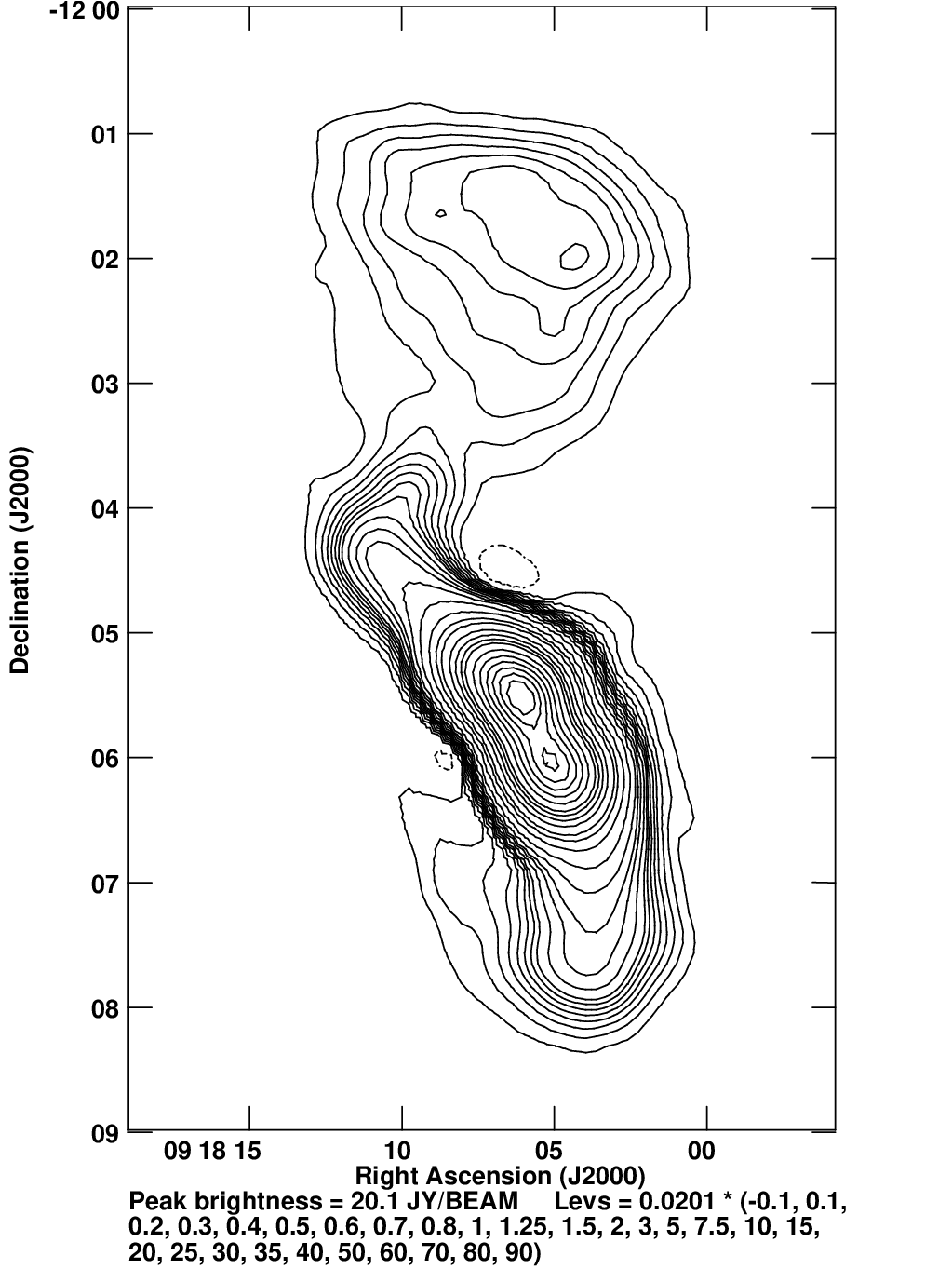}
\includegraphics[width=3.25in]{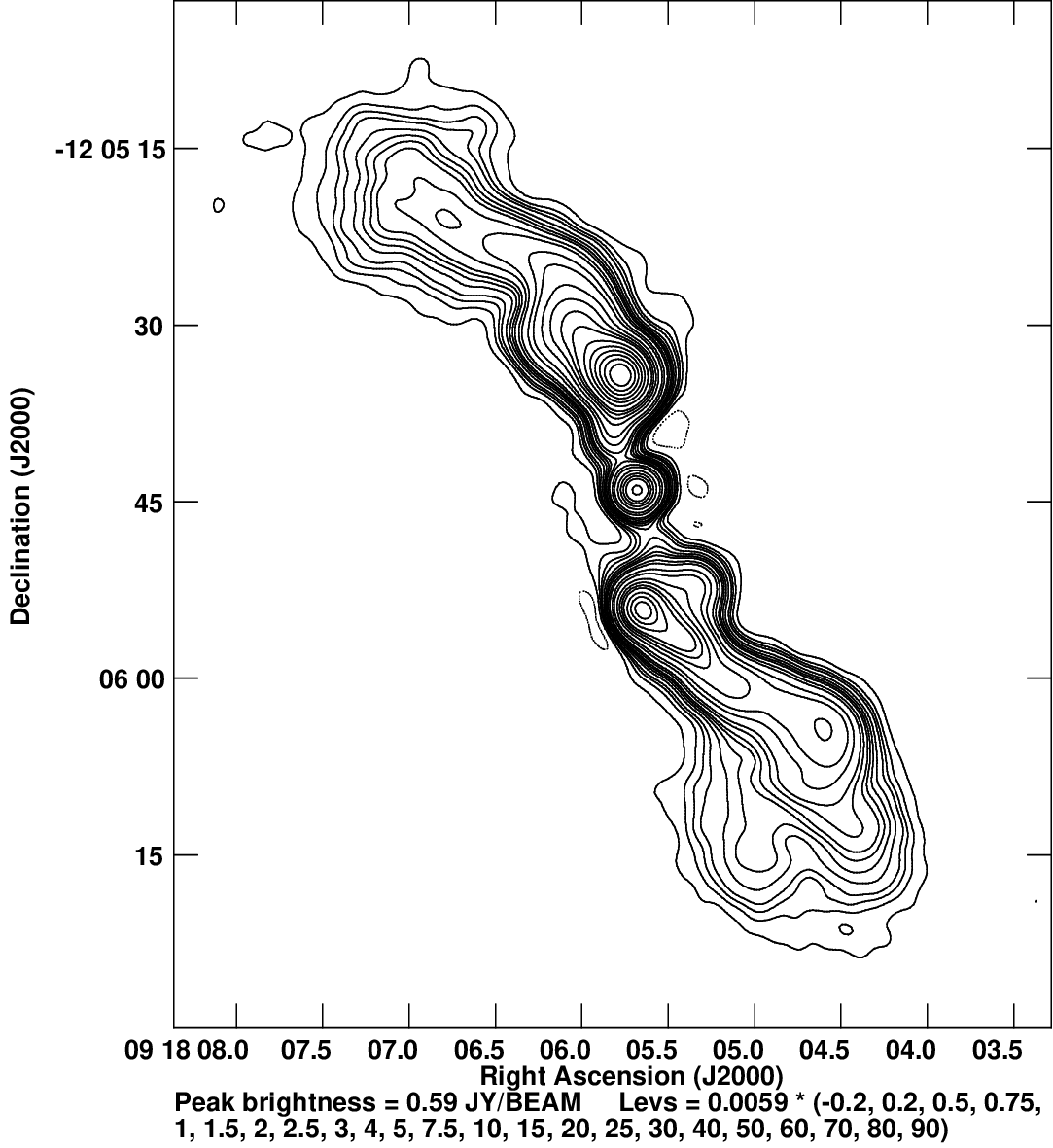}
}}
\centerline{\parbox{6.5in}{
    \caption{\small (Left) 3C218 at 1040 MHz with 28 arcsecond
      resolution.  (Right) The central regions of 3C218 at 11.06 GHz,
      with 2.65 arcsecond resolution.  At this frequency the extended
      emission has faded below detectability.  }
\label{fig:HydraA}}}
\end{figure}

Hydra A (3C218) is a FRI radio galaxy lying in the center of the Abell
cluster A780.  It has extensive, low-brightness large-scale structure
as well as compact jet and nuclear structure near the center.  These
features are illustrated in Figure~\ref{fig:HydraA}.  VLA images taken
at 74 MHz by \citet{Lane04} show the southern low-brightness emission
extending much further to the east.  The overall extent of about 8
arcminutes, complexity of structure, and the relatively strong nuclear
core make this a problematic object for flux density scale calibration
above 1 GHz.  Its spectrum is shown in Fig~\ref{fig:FourPlotE}.

\begin{figure}[!ht]
\centerline{\hbox{
\includegraphics[width=3.25in]{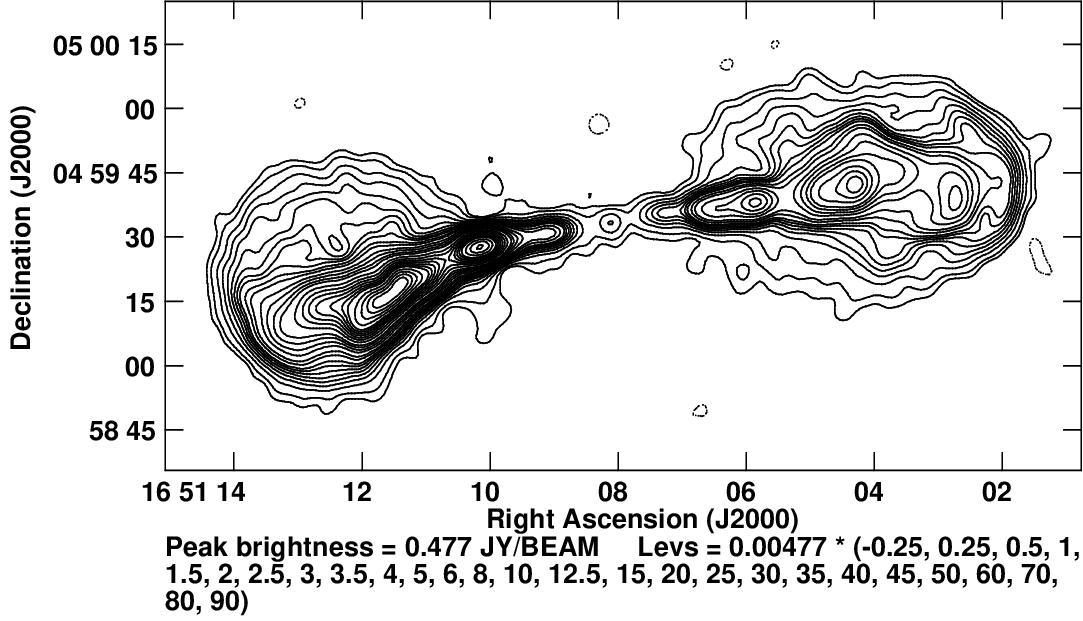}
\includegraphics[width=3.25in]{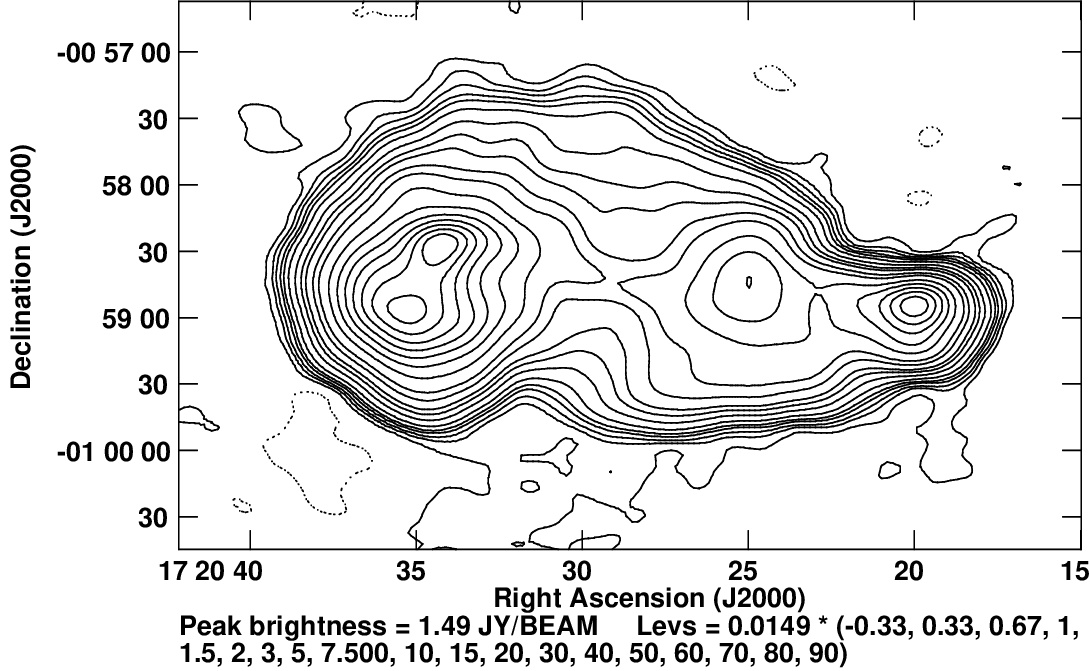}
}}
\centerline{\parbox{6.5in}{
\caption{\small (Right)  Hercules A at 6564 MHz with 5 arcseconds
  resolution.  (Left) 3C353 at 2948 MHz and 13 x 25 arcsecond
  resolution. }
\label{fig:Herc3C353}}}
\end{figure}
\subsection{Hercules A}
Hercules A is a radio galaxy with redshift $z=0.155$.  Its structure
at 6564 MHz with 5 arcseconds resolution is shown in
Figure~\ref{fig:Herc3C353}.  The maximum extent of 3.1 arcminutes
makes this a difficult object for calibration by high resolution
arrays, although its very weak nuclear emission strongly suggests the
source flux density will be very stable.  Its spectrum is shown in
Fig~\ref{fig:FourPlotE}.
\subsection{3C353}
3C353 is a nearby radio galaxy of redshift $z=0.0304$, the radio structure is
very large (5.3 arcminutes), and quite complicated.  An image at 2948
MHz, with 25 x 13 arcsecond resolution, is shown in
Figure~\ref{fig:Herc3C353}.  As the nuclear
flux is very weak (less than 0.1 Jy, so much less than 0.1\% of the
total flux), the total flux density is expected to be very stable.  
Its spectrum is shown in Fig~\ref{fig:FourPlotE}.

\section{Discussion}

What is the best means for providing accurate gain calibration for an
interometric array?  In principle, accurate amplitude calibration can
be done without the use of an external standard.  As shown in PB13,
what is required is good knowledge of the antenna gain (or aperture
efficiency) and of the on-board noise calibration power.  These
quantities are functions of antenna elevation and observing frequency,
and will differ, possibly significantly, amongst the antennas of an
array. It has generally been argued that the effort involved in
measuring, monitoring, and implementing the necessary parameters for
an array is not cost-effective, and that external calibration schemes
are sufficient.  Additionally, it should be added that external
factors, such as weather, are not accounted for with an internal-only
calibration scheme.

Because of these issues, most arrays utilize an external flux density
standard source for amplitude gain calibration.  Provided that antenna
and system electronics gains do not change between observations of
the target and the standard gain calibrator, accurate gain calibration
requires only application of correlation ratios between the target and
reference objects.  

The question then becomes: What makes a good flux density scale
calibrator?  It is easy to list the ideal properites -- unresolved at
all bands and baselines, unchanging on timescales of decades, and
strong enough so that system noise and background source confusion are
negligible.  There is no such source.  It seems to be a law of nature
that sources small enough to remain unresolved over a wide range of
resolutions are necessarily variable.  Furthermore, the strongest
stable sources are nearly all significantly resolved to modern
high-resolution arrays.

Thus, we must utilize partially-resolved objects for amplitude gain
calibration.  What criteria should we apply when deciding which
sources to utilize?  These include:
\begin{itemize}
\item Stable for long periods -- preferably decades, but at least
  years.  Less satisfactory, but potentially acceptable, is a known
  secular change.  Although slowly variable objects can be monitored
  fairly easily, the overhead in doing so and disseminating results
  imposes significant additional costs and risks.
\item Small enough not to be resolved out by the longest spacings in a
  given array.  Use of a partially resolved model requires an accurate
  model as a function of frequency, but if the source is non-variable,
  there is no fundamental problem with this.  Additionally, the object
  should be much smaller than the component antennas' primary beam, so
  that frequency-dependent corrections are not necessary.  
\item Strong enough so the effects of confusion and system noise are
  neglible in the solution for the antenna gains.  The former problem
  is an issue for lower frequencies, the latter for higher
  frequencies.
\end{itemize}

We have reviewed the characteristics of the 20 sources included in
this study, and generated criteria to judge the suitability of each
for calibration by arrays of the scale of the VLA, over the 50 MHz --
50 GHz span of this study.  Specific criteria, and comments, are given
below. We note that many of the criteria chosen are arbitrary, and
many of the ranges given are specific to the VLA, and its 25-meter
primary antennas.  It should be relatively straightforward to modify
these for any given array.  
\begin{itemize}
\item {\bf Stability:} The total flux density should not change by
  more than 2\% per year, unless in a preditable way.  For extended
  sources comprising a (presumably variable) nuclear core, if we
  assume the core can change 40\% in one year (a likely worst case),
  the nucleus flux must be less than 5\% of the total.
\item {\bf Angular Size:}  The source should not be attenuated by the
  primary beam by more the 2\%.  This criterion is strongly
  frequency-dependent, both because of diffraction-effects for the
  primary beam, and from spectral index gradients in FR1-type radio
  galaxies.    
\item {\bf Resolution:} The visibility of the longest spacing should
  be at least 5\% of the total, to ensure that a source model predicts
  enough visibility for stable gain solutions for the most distant
  antennas.  Essentially, this requirement tries to limit the
  diffusiveness and complexity of the necessary source model.
  Furthermore, the visibility amplitude on the longest spacing must be
  at least 10 times the rms noise in the visibilities for the time and
  frequency averaging used in the solution.  For the VLA, this
  translates to $\sim$ 100 mJy for the central bands (L through X),
  rising to 300-500 mJy at both low and high frequencies.  
\item {\bf Background Source Confusion} Nearby confusing sources are
  effectively a source of noise in the gain calibration.  Modelling
  these is difficult, since their effect on the visibilities will
  depend on the time and frequency averaging used in the calibration
  solution.  To minimize gain solution variations, we require the
  confusion noise in the visibilities to be less than 50\% of the
  total flux, and must resolve out before 1/3 of the maximum baseline.
\end{itemize}
The criterion regarding confusion might be considered too loose, but
in fact the contributing background sources have random phases, such
that the antenna-based gain solutions have remarkable isolation to
their presence.  In fact, they can be considered as an additional
source of noise.  The angular size criterion is not critical -- if the
primary beam shape is known, the attenuation can be calculated since
the structure is known.  However, this is another model-dependent
nuisance which is best avoided.  

These points are illustrated in the following figures.
Figure~\ref{fig:Confusion} shows the effect of background source
confusion of the visibilities for 3C138 at 322 MHz.  For this source,
the rise is visibility amplitudes at less than 500 wavelengths
baseline is due to the Crab Nebula (3C144), located 6.4 degrees
away. The reduction in the confusing visibility is a combination of
the resolution of the source, combined with both bandwidth and time
averaging of the visibilities.  
\begin{figure}[!ht]
\centerline{\hbox{
\includegraphics[width=3.25in]{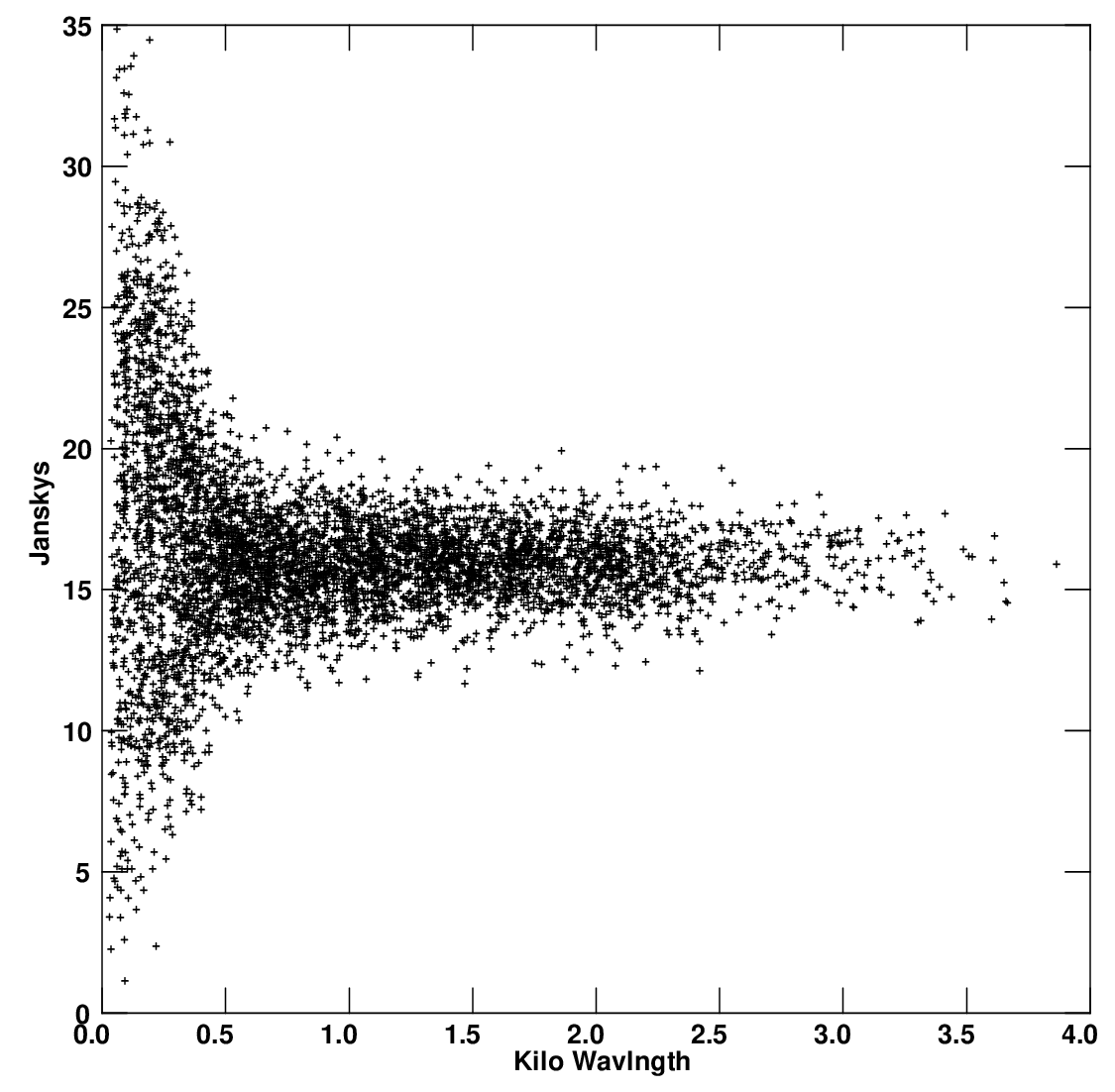}
\includegraphics[width=3.25in]{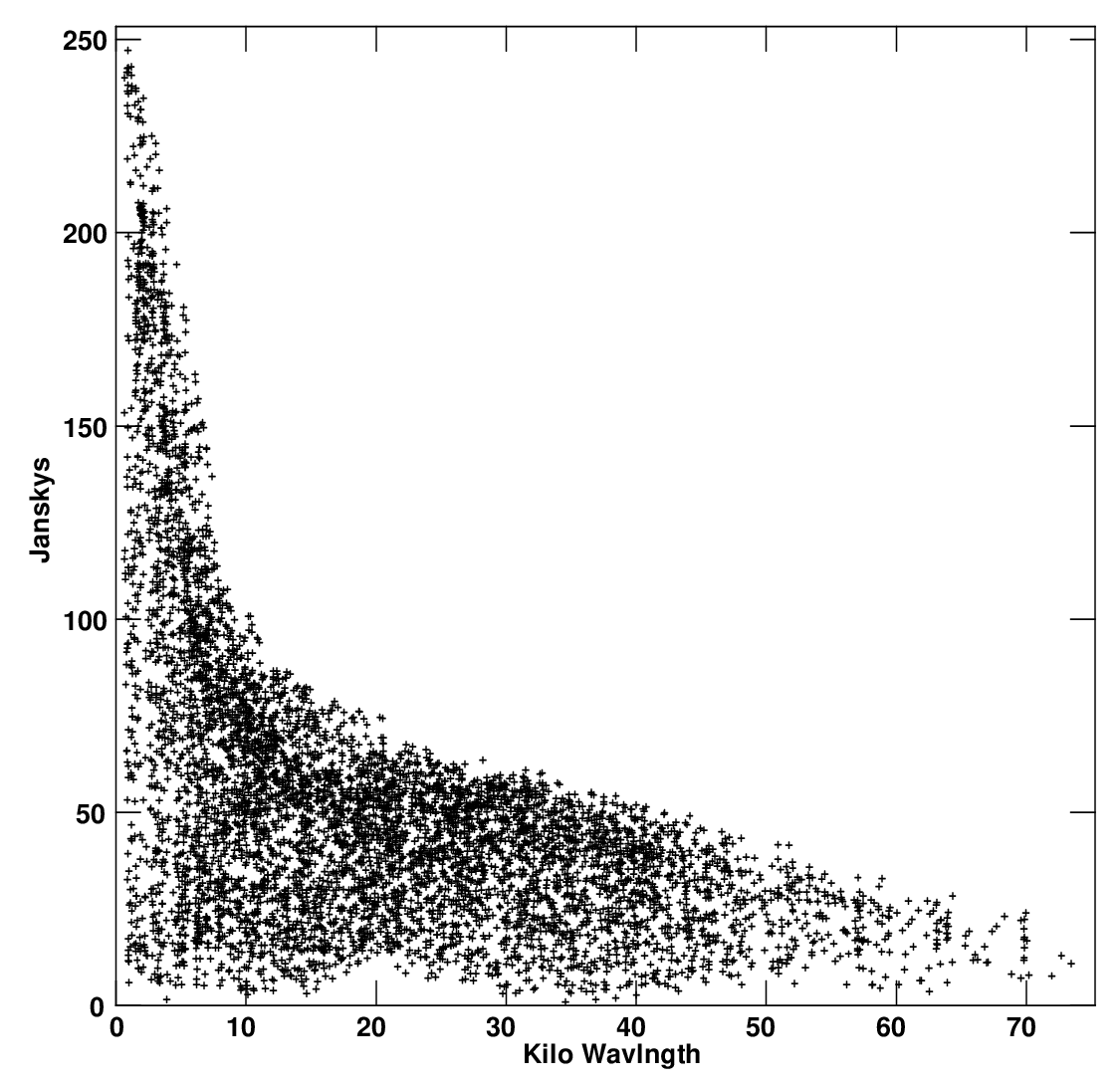}}}
\centerline{\parbox{6.5in}{
    \caption{\small (Left) Visibility function of 3C138 at 344
      MHz. The scatter is the visibility amplitudes is due to the Crab
      Nebula, located 6.4 degrees away.  
      (Right) Cygnus A at 6564 MHz. The smooth lobes are resolved out
      at about $10^4$ wavelengths baseline.   The slow decline
      thereafter is from the gradual resolution of the bright
      hotspots.   }
\label{fig:Confusion}}}
\end{figure}
At low frequencies, the confusion from 3C144 makes the short spacings
unuseable for calibration.  Interferometers shorter than
$\sim$1K$\lambda$ (e.g. the VLA in D configuration) should not use
this source.  The plot shown, taken from C configuration data,
indicates that calibration can be accomplished provided the short
spacings are excluded.  

The right panel shows the visibility of Cygnus A at 6564 MHz.  The
source is heavily resolved (and is also large enough that it fails the
primary beam resolution criterion), but the longest spacings retain
enough visibility that a viable model can be utilized for resolutions
at least as good as 5 arcseconds.  

\begin{figure}[!ht]
\centerline{\hbox{
\includegraphics[width=3.25in]{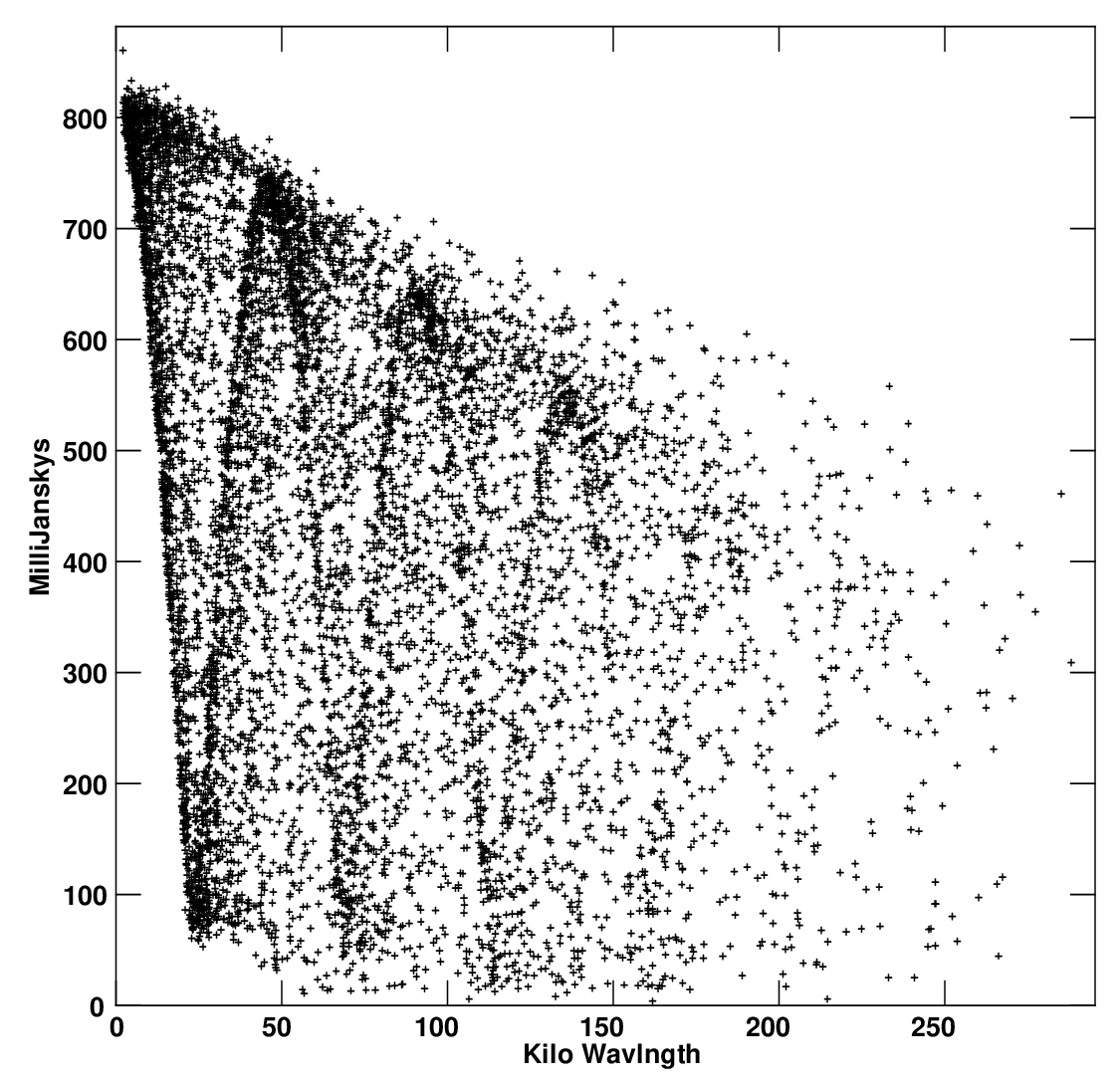}
\includegraphics[width=3.25in]{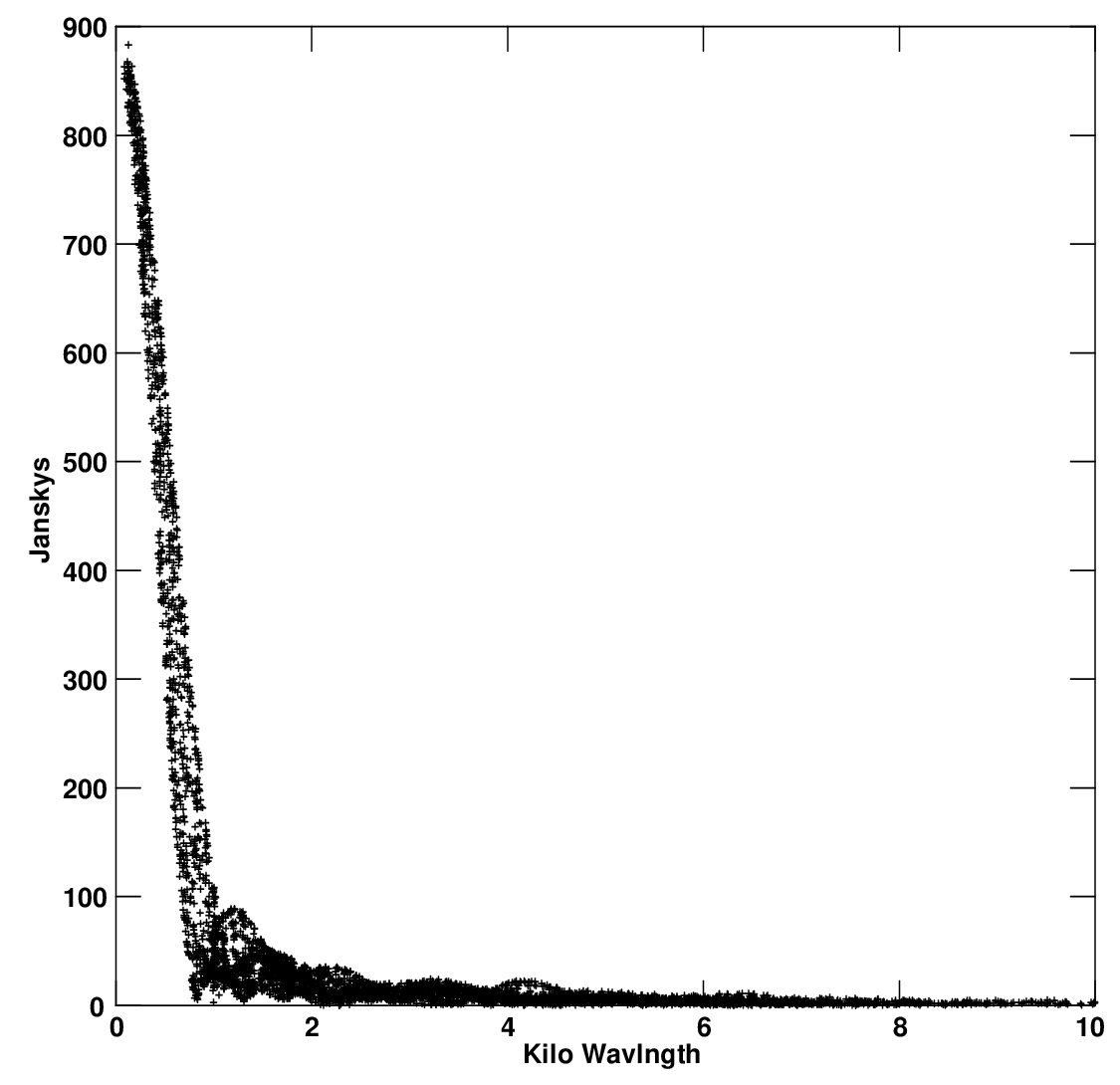}
}}
\centerline{\parbox{6.5in}{
    \caption{\small (Left) Visibility function of 3C295 at 25564 MHz.
      This clearly visible beating is from the two bright hotspots,
      separted by 5 arcseconds.  (Right) 3C144 at 1040 MHz.  The rapid
      drop to very low visibilities is due to the smooth nebular
      structure, with no strong gradients. }
\label{fig:Resolution}}}
\end{figure}

Figure~\ref{fig:Resolution} shows two sources with significant, but
very different resolution effects -- 3C295 and 3C144.

The left panel shows 3C295 at 25564 MHz.  Although the source
structure comprises two lobes separated by 5 arcseconds, the
individual lobes are sufficiently small that there is enough
visibility on the longest spacings that a stable solution with
adequate SNR can be expected for intereferometers with arcsecond
resolution.  The right panel shows 3C144 at 1040 MHz, with maximum 3
Km baselines.  Although the longest spacings see about 1 Jy flux, this
is too small a fraction of the total flux of 900 Jy for a stable
solution to be expected.  

A summary of the viability of each of the 20 sources for VLA
observations is given in Table~\ref{tab:CalSuit}.  Note that as many
of the criteria are strongly frequency dependent, we have utilized a
frequency of 1 GHz.  In such cases, the notes column contains
additional information.  

\begin{deluxetable}{lllll}
\tablecaption{Flux Calibrator Suitability for VLA
\label{tab:CalSuit}}
\tablewidth{0pt}
\tablehead{
\colhead{Source}&\colhead{Var\tablenotemark{a}}&
\colhead{MaxFreq\tablenotemark{b}}&\colhead{MinVis\tablenotemark{c}}&
\colhead{Comments}}
\startdata
J0133-3629&OK     &0.7&0 -- 10& Extremely large and diffuse\\
3C48      &Irreg. &All&0 -- 5000&Can be used up to 40 GHz for VLA\\
Fornax A  &OK     &0.2&0 -- 0.05&Generally unsuited for interferometers\\
3C123     &OK     &15 &0 -- 1000&\\
J0444-2809&OK     &5  &0 -- $>$50&Probably useful to longer spacings\\
Pictor A  &$<$4     &1.2&0 -- 30&\\
3C138     &Irreg. &All&0 -- 5000&Strongly confused below 500$\lambda$ at P-band\\
3C144     &Slow   &1.2&0 -- 2&Too diffuse for use at high frequencies\\
3C147     &Irreg. &All&0 -- 5000&\\
3C196     &OK     &All&0 -- 1000&\\
3C218     &$<$10    &1.5&0 -- 250&\\
3C274     &$<$4     &0.7&0 -- 10 &Compact structure could be used at high resn.\\
3C286     &OK     &All&0 -- 5000&\\
3C295     &OK     &All&0 -- 2000&Too weak at high frequencies and resn.\\
3C348     &OK     &3  &0 -- 100 &\\
3C353     &OK     &2  &0 -- 20  &\\
3C380     &$<$1?    &30 &0 -- 5000&Nuclear flux below 1 GHz uncertain\\
3C405     &OK     &5  &0 -- 250&\\
3C444     &OK     &5  &0 -- 20&\\
3C461     &Slow   &1  &0 -- 5&\\
\enddata
\tablenotetext{a}{Variability Criterion -- maximum frequency in GHz
  for sources with strong nuclei}
\tablenotetext{b}{Frequency in GHz at which primary beam resolution criterion met}
\tablenotetext{c}{Baseline range (kilo-wavelengths) to meet confusion,
  structure and sensitivity criteria.}
\end{deluxetable}

\section{Conclusions}

We have defined a comprehensive new flux density calibration scale for
radio astronomy, valid between 50 MHz and 50 GHz.  Polynomial
coefficients for 20 proposed calibrators, distributed over both
hemispheres, and useable both for single dishes and for
interferometers of up to $\sim$5000K$\lambda$ baseline length, are
given.  The majority of the sources are stable over long periods of
time.  Some are slowly time variable, and will need regular monitoring
to be useful for accurate flux density calibration.

This scale replaces that proposed by us in 2013, as it extends that
scale downwards from 1 GHz to 50 MHz.  The new scale is identical (to
the quoted errors) to the old scale above 2 GHz.  Correction factors
for the older Baars et al. scale and the Scaife and Heald scale are
given.  

The chief weakness of our new scale is at frequencies below 240 MHz.
The polynomical expressions are entirely dependent on a single
measurement made with the VLA's `legacy' 74 MHz system for 13 of our
20 sources.  For the remaining seven sources, there is no VLA
measurement, so our expressions cannot be used below 200 MHz.  While
we have no reason to doubt the accuracy of the old measurements,
confidence would be increased when data from the VLA's new low
frequency system are available.

Probably most useful for confirming the accuracy of our proposed scale
would be measurements made by the new generation of low-frequency
facilities, notably the MWA and LOFAR.  These would fill the gaps
below 250 MHz for our measurements, confirm the ratios that we have
determined, and fill in the large gap between 74 and 240 MHz.  

A final point worthy of mention is the question of `what calibrates
the calibrator'?  Out scale at low frequencies is entirely based on
observations of Cygnus A with absolutely calibrated antennas and
interferometers, nearly all of these done more than 40 years ago.
While we do not doubt the accuracy of these efforts, the availability
of modern technologies suggests that more accurate and robust
measurements of this fundamental standard should be possible today.
Given that the errors of our scale are entirely due to the error in
the primary calibrators, better accuracy can only be obtained with
better fundamental standards.


\begin{thebibliography}{}
\bibitem[Baars and Hartsuijker (1972)]{BH72} Baars, J.W.M., and
  Hartsuijker, A.P., 1983 \aap, 17, 172
\bibitem[Baars, Mezger, and Wendker (1965)]{BMW65} Baars, J.W.M.,
  Mezger, P.G., and Wendker, H., 1965,\apj, 142, 122
\bibitem[Baars et al.(1977)]{Baa77} Baars, J.W.M., Genzel, R.,
  Paulinty-Toth, I.K.K., and Witzel, A. 1977, \aap, 61, 99
\bibitem[Beitenholz et al. (1991)]{B91} Beitenholz, M.B., Kronberg,
  P.P., Hogg, D.E., and Wilson, A.S. \apjl, 373, L59
\bibitem[Conway, Kellermann, and Long (1963)]{CKL63} Conway,
  R.G., Kellermann, K.I., and Long, R.J., 1963, \mnras, 125, 261
\bibitem[Kassim et al. (2007)]{K07} Kassim, N.E., Lazio, T.J.W.,
  Erickson, W.E., Perley, R.A., Cotton, W.D., Greisen, E.W., Cohen,
  A.S., Hicks, B., Schmitt, H.R. and Katz, D. 2007, \apjs, 172, 686
\bibitem[Kellermann (1964)]{K64} Kellermann, K.I.,  1964, \aj, 69, 205
\bibitem[Lane et al. (2004)]{Lane04} Lane, W.M., Clarke, T.E., Taylor,
  G.B., Perley, R.A., and Kassim, N.E. \ 2004, \aj, 127, 48
\bibitem[Lister et al. (2013)]{List13} Lister, M.L., Aller, M.F., Aller, H.D.,
    Homan, D.C., Kellermann, K.I., Kovaley, Y.Y., Pushkarev, A.B.,
    Richards, J.L., Ros, E., and Savolainen, T. 2013, \aj, 146, 120
\bibitem[Pacholczyk (1970)]{Pach70} Pacholczyk, A.G. 1970, Radio
  Astrophysics (W.H Freeman and Company)
\bibitem[Partridge et al. (2016)]{Par16} Partridge, B., Lopez-Caniego,
  M., Perley, R.A., Stevens, J., Butler, B.J., Rocha, G., Walter, B.,
  and Zacchei, A. , 2016, \apj, 821, 61
\bibitem[Perley and Butler (2013)]{PB13} Perley, R.A., and Butler,
  B.J. 2014,  \apjs, 204, 19
\bibitem[Perley (2016)]{P16} Perley, Rick,  2016, EVLA Memo 195
\bibitem[Roger, Bridle, and Costain (1973)]{RBC73} Roger, R.S.,
  Bridle, A.H., and Costain, C.H. 1973, \aj, 78, 1030
\bibitem[Scaife and Heald (2012)]{SH12} Scaife, A.M.M., and Heald,
  G.R. 2012, \mnras, 423, L30
\bibitem[Wills (1973)]{W73} Wills, B.J. 1973, \apj, 180, 335
\end{thebibliography}
\end{document}